\documentclass[traditabstract]{aa}
\pdfoutput=1
\usepackage{verbatim} 
\usepackage{amsmath}
\usepackage{amssymb}
\usepackage{graphicx}
\usepackage{booktabs}
\usepackage[authoryear]{natbib}
\usepackage[colorlinks=true,linkcolor=blue,urlcolor=blue,citecolor=blue]{hyperref}
\usepackage{xspace}

\providecommand{\tabularnewline}{\\}

\usepackage{txfonts}
\usepackage{twoopt}
\bibpunct{(}{)}{;}{a}{}{,} 
\newcommand{\msun}{{\rm M}_{\odot}}
\newcommand{\lsun}{{\rm L}_{\odot}}
\newcommand{\rsun}{{\rm R}_{\odot}}

\newcommand{\bonnsai}{\mbox{\textsc{Bonnsai}}\xspace}
\titlerunning{The \bonnsai project}
\authorrunning{Fabian~R.N.~Schneider~et~al.}

\begin{document}

\title{\bonnsai: a Bayesian tool for comparing stars with stellar evolution models\thanks{\bonnsai is available
through a web-interface at\newline\href{http://www.astro.uni-bonn.de/stars/bonnsai}{http://www.astro.uni-bonn.de/stars/bonnsai}.}}

\author{F.R.N.~Schneider\inst{\ref{AIFA}}\thanks{fschneid@astro.uni-bonn.de}
\and N.~Langer\inst{\ref{AIFA}}
\and A.~de Koter\inst{\ref{AMSTERDAM},\ref{LEUVEN}}
\and I.~Brott\inst{\ref{VIENNA}}
\and R.G.~Izzard\inst{\ref{AIFA}} 
\and H.H.B.~Lau\inst{\ref{AIFA}}}

\institute{Argelander-Institut f{\"u}r Astronomie der Universit{\"a}t Bonn, Auf dem H{\"u}gel~71, 53121~Bonn, Germany\label{AIFA} 
\and Astronomical Institute 'Anton Pannekoek', Amsterdam University, Science Park 904, 1098 XH, Amsterdam, The Netherlands\label{AMSTERDAM} 
\and Instituut voor Sterrenkunde, KU Leuven, Celestijnenlaan 200D, 3001, Leuven, Belgium\label{LEUVEN} 
\and University of Vienna, Department of Astronomy, T{\"u}rkenschanzstr. 17, 1180 Vienna, Austria\label{VIENNA}}

\date{Received day month 2014 / Accepted day month 2014}

\abstract{Powerful telescopes equipped with multi-fibre or integral field spectrographs
combined with detailed models
of stellar atmospheres and automated fitting techniques allow for
the analysis of large number of stars. These datasets contain a wealth
of information that require new analysis techniques to bridge the
gap between observations and stellar evolution models. To that end,
we develop \bonnsai (BONN Stellar Astrophysics Interface), a Bayesian
statistical method, that is capable of comparing all available observables
simultaneously to stellar models while taking observed uncertainties
and prior knowledge such as initial mass functions and distributions
of stellar rotational velocities into account. \bonnsai can be used
to (1) determine probability distributions of fundamental stellar
parameters such as initial masses and stellar ages from complex datasets, (2) predict stellar parameters
that were not yet observationally determined and (3) test stellar
models to further advance our understanding of stellar evolution.
An important aspect of \bonnsai is that it singles out stars that cannot
be reproduced by stellar models through $\chi^{2}$ hypothesis tests
and posterior predictive checks. \bonnsai can be used with any set
of stellar models and currently supports massive main-sequence single
star models of Milky Way and Large and Small Magellanic Cloud composition.
We apply our new method to mock stars to demonstrate its functionality
and capabilities. In a first application, we use \bonnsai to test the
stellar models of Brott et al. (2011a) by comparing the stellar ages
inferred for the primary and secondary stars of eclipsing Milky Way
binaries of which the components range in mass between $4.5$ and
$28\,\msun$. Ages are determined from dynamical masses and radii
that are known to better than 3\%. We show that the stellar models
must include rotation because stellar radii can be increased by several
percent via centrifugal forces. We find that the average age
difference between the primary and secondary stars of the binaries
is $0.9\pm2.3\,\mathrm{Myr}$ (95\% CI), i.e. that the stellar models
reproduce the Milky Way binaries well. The predicted effective temperatures
are in agreement for observed effective temperatures for stars cooler than $25,000\,\mathrm{K}$.
In hotter stars, i.e. stars earlier than B1--2V and more massive than
about $10\,\msun$, we find that the observed effective temperatures
are on average hotter by $1.1\pm0.3\,\mathrm{kK}$ (95\% CI) than
predicted by the stellar models. The hotter temperatures consequently
result in bolometric luminosities that are larger by $0.06\pm0.02\,\mathrm{dex}$
(95\% CI) than those predicted by the models.}


\maketitle

\section{Introduction}\label{sec:introduction}

Stars are the building blocks of galaxies and hence the Universe.
Our knowledge of their evolution is deduced from detailed comparisons
of observations to theoretical models. The interface, where observations
meet theory, is often provided by the Hertzsprung-Russell (HR) diagram
and its relative, the colour-magnitude diagram. In an HR diagram,
two surface properties of stars, the effective temperature and the
luminosity, are compared to stellar evolutionary models to, e.g.,
determine fundamental stellar parameters like initial mass and age
that are both inaccessible by observations alone and of utmost importance
to modern astrophysics. 

With the advent of large spectroscopic surveys\footnote{Sloan Digital Sky Survey (SDSS), Sloan Extension for Galactic 
Understanding and Exploration (SEGUE/SDSS), Galactic O Star Spectroscopic Survey (GOSSS),
RAdial Velocity Experiment (RAVE), Spectroscopic Survey of Galactic O and WN stars (OWN),
Instituto de Astrof{\'i}sica de Canarias OB star survey (IACOB)}
such as the Gaia-ESO survey \citep{2012Msngr.147...25G}, SEGUE/SDSS \citep{2009AJ....137.4377Y},
GOSSS \citep{2011hsa6.conf..467M,2013msao.confE.198M}, RAVE \citep{2006AJ....132.1645S},
OWN \citep{2010RMxAC..38...30B}, IACOB \citep{2011BSRSL..80..514S,2014A&A...562A.135S},
the VLT FLAMES Survey of Massive Stars \citep{2005A&A...437..467E,2006A&A...456..623E}
and the VLT-FLAMES Tarantula Survey \citep{2011A&A...530A.108E},
much more is known about stars than just their positions in the HR
diagram: surface abundances, surface gravities, surface rotational
velocities and more. Such a wealth of information not only allows
the determination of fundamental stellar parameters with high precision
but also to thoroughly test and thus advance our understanding of
stellar evolution. However, bridging the gap between such manifold
observations and stellar models requires more sophisticated analysis
techniques than simply comparing stars to models in the HR diagram
because the comparison needs to be performed in a multidimensional
space. 

Bayesian inference is widely applied in determining cosmological parameters
and offers a promising framework to infer stellar parameters from
observations. Steps into this direction have been taken by \citet{2004MNRAS.351..487P}
and \citet{2005A&A...436..127J} to infer stellar ages and masses
from colour-magnitude diagrams. Since then, Bayesian modelling has
been used by several authors to infer a wide range of stellar parameters
from spectroscopy, photometry, astrometry, spectropolarimetry and
also asteroseismology \citep[e.g.][]{2006A&A...458..609D,2007ApJS..168..297T,2007MNRAS.377..120S,2010MNRAS.407..339B,2011A&A...530A.138C,2012ApJ...749..109G,2012MNRAS.420..773P,2012MNRAS.427.1847B,2013MNRAS.429.3645S,Schoenrich01092014}.
Bayesian inference is further used to derive properties of stellar 
ensembles such as cluster ages, mass functions
and star formation histories \citep[e.g.][]{2006ApJ...645.1436V,2009ApJ...696...12D,2009AnApS...3..117V,2013ApJ...762..123W,2013MNRAS.435.2171W,2014arXiv1405.3287D}.
The big advantage of a Bayesian approach is the knowledge of full
probability distribution functions of stellar parameters that take
observational uncertainties and prior knowledge into account.

Incorporating prior knowledge may be essential because the determination
of stellar parameters can otherwise be biased \citep[e.g.][]{2004MNRAS.351..487P}.
For example, the evolution of stars speeds up with time such that
stars spend different amounts of time in various evolutionary phases
(this even holds for stars on the main-sequence). This knowledge is
typically neglected when determining stellar parameters from the position
of stars in an HR diagram and may thus result in biased stellar parameters. 

In this paper, we present a method that \emph{simultaneously} takes
all available observables, their uncertainties and prior knowledge
into account to determine stellar parameters and their full probability
density distributions from a set of stellar models. This method further
allows us to predict yet unobserved stellar properties like rotation
rates or surface abundances. By applying sophisticated goodness-of-fit
tests within our Bayesian framework, we are able to securely identify
stars that cannot be reproduced by the chosen stellar models which
will help to improve our understanding of stellar evolution. 

\begin{table}
\caption{Initial mass ranges $M_{\mathrm{ini}}$, age ranges and initial rotational
velocity ranges $v_{\mathrm{ini}}$ of the Bonn Milky Way (MW), Large
Magellanic Cloud (LMC) and Small Magellanic Cloud (SMC) stellar models
\citep{2011A&A...530A.115B,Koehler+2014}. The models contain main-sequence
single stars. We discuss the metallicities of the models in Sec.~\ref{sec:comparison-of-ages}.}

\begin{centering}
\begin{tabular}{cccc}
\toprule 
Stellar models & $M_{\mathrm{ini}}/\msun$ & Age / Myr & $v_{\mathrm{ini}}/\mathrm{km}\,\mathrm{s^{-1}}$\tabularnewline
\midrule
\midrule 
Bonn MW & 5--50 & 0--100 & 0--600\tabularnewline
Bonn LMC & 5--500 & 0--100 & 0--600\tabularnewline
Bonn SMC & 5--60 & 0--100 & 0--600\tabularnewline
\bottomrule
\end{tabular}
\par\end{centering}

\label{tab:mass-and-velocity-ranges}
\end{table}

Currently, \bonnsai supports the Bonn stellar models for the Milky
Way (MW), Large Magellanic Cloud (LMC) and Small Magellanic Cloud
\citep[SMC;][]{2011A&A...530A.115B,Koehler+2014} with initial mass
and rotational velocity ranges given in Table~\ref{tab:mass-and-velocity-ranges}.
We plan to integrate further stellar model grids in the future
and make \bonnsai available through a web-interface\footnote{\href{http://www.astro.uni-bonn.de/stars/bonnsai}{http://www.astro.uni-bonn.de/stars/bonnsai}}.

We describe the \bonnsai approach in Sec.~\ref{sec:method} and apply
it to mock stars in Sec.~\ref{sec:mock-stars} to show its functionality
and to demonstrate its capabilities. In Sec.~\ref{sec:real-data}
we apply \bonnsai to high-precision observations of Milky Way binaries
to thoroughly test the Milky Way stellar models of \citet{2011A&A...530A.115B}
and conclude in Sec.~\ref{sec:conclusions}.

\section{Method}\label{sec:method}

Besides the observables, there are two main ingredients in \bonnsai.
On the one hand, there is the statistical method and on the other
hand there are the stellar models. The statistical method uses the
stellar models to derive stellar parameters from a given set of observables.
In the following, we describe our statistical method (Secs.~\ref{sec:bayes-theorem}--\ref{sec:priors}
and~\ref{sec:goodness-of-fit}--\ref{sec:bonnsai-in-practice}) and
the so far implemented stellar models (Sec.~\ref{sec:stellar-model-grids}).

\subsection{Bayes' theorem}\label{sec:bayes-theorem}

Bayes' theorem directly follows from the definition of conditional
probabilities. Let $P(M|D)$ be the conditional probability that an
event $M$ occurs given that an event $D$ already took place, i.e.
$P(M|D)=P(M\cap D)/P(D)$ where $P(M\cap D)$ is the joint probability
of both events and $P(D)\neq0$ the probability for the occurrence
of event $D$. We then arrive at Bayes' theorem,
\begin{equation}
P(M|D)=\frac{\frac{P(M\cap D)}{P(M)}P(M)}{P(D)}=\frac{P(D|M)P(M)}{P(D)}.\label{eq:bayes-theorem}
\end{equation}
In the context of Bayesian inference, $M$ is the model parameter,
$D$ is the (observed) data and
\begin{itemize}
\item $P(M|D)$ is the \emph{posterior} probability function, i.e. the probability
of the model parameter given the data,
\item $P(D|M)$ is the \emph{likelihood}, i.e. the probability of the data
given the model parameter,
\item $P(M)$ is the \emph{prior} function, i.e. the probability of the
model parameter,
\item and $P(D)$ is the \emph{evidence} that serves as a normalisation
constant in this context because it does not depend on the model parameter.
\end{itemize}
The likelihood function per se is \emph{not a} probability distribution
of the model parameters that we seek to derive but it describes how
likely each value of an observable is given a model. To derive the
probability distribution of the model parameters, i.e. the posterior
probability distribution, we apply Bayes' theorem. In case of a uniform
prior function, $P(M)=\mathrm{const.}$, the likelihood is the posterior
probability distribution and Bayesian inference reduces to a maximum
likelihood approach.

\subsection{Bayesian stellar parameter determination}\label{sec:bayseian-parameter-determination}

In the case of stellar parameter determination, the data $\vec{d}$
are now an $n_{\mathrm{obs}}$-dimensional vector containing the $n_{\mathrm{obs}}$
observables like luminosities $L$, effective temperatures $T_{\mathrm{eff}}$,
surface abundances etc.\footnote{What we call observables are actually 
not observables but quantities derived from observations, e.g.\ from modelling 
of stellar spectra. For clarity we nevertheless 
use the phrase observables in this paper.} The model parameter $\vec{m}$ for rotating
single stars is a 4-dimensional vector consisting of the initial mass
$M_{\mathrm{ini}}$, the initial rotational velocity at the equator
$v_{\mathrm{ini}}$, the initial chemical composition/metallicity
$Z$ and the stellar age $\tau$. Further physics that alters the
evolution of stars, like magnetic fields or duplicity, may be added
to the stellar models and hence to the model parameters if needed.
The model parameters uniquely map to the observables; e.g. stellar
models predict luminosities, effective temperatures, etc. for given
stellar initial conditions and age, $\vec{m}=(M_{\mathrm{ini}},v_{\mathrm{ini}},Z,\tau)\rightarrow\vec{d}(\vec{m})=(L,T_{\mathrm{eff}},\ldots)$.
The reverse mapping is not always unique as evident from Fig.~\ref{fig:mock-stars-hrd}
where stellar tracks of different stars cross each other in the HRD,
i.e. different model parameters, $\vec{m}$, can predict the same
observables $\vec{d}$. 

Bayes' theorem in terms of probability \emph{density} functions reads, analogously
to Eq.~\ref{eq:bayes-theorem},
\begin{equation}
p(\vec{m}|\vec{d})\propto p(\vec{d}|\vec{m})p(\vec{m}).\label{eq:our-bayes-theorem}
\end{equation}
The proportionality constant follows from the normalisation condition,
\begin{equation}
\int_{\vec{m}}\, p(\vec{m}|\vec{d})\,\mathrm{d}\vec{m}=1,\label{eq:normalisation}
\end{equation}
where $\mathrm{d}\vec{m}=\mathrm{d}M_{\mathrm{ini}}\,\mathrm{d}v_{\mathrm{ini}}\,\mathrm{d}Z\,\mathrm{d}\tau$.
To compute $p(\vec{m}|\vec{d})$ we must define the likelihood and
prior functions, which we do in the following two sections.

\subsection{Likelihood function}\label{sec:likelihood}

We assume two-piece Gaussian likelihood functions to compute the posterior
probability distribution according to Bayes' theorem, Eq.~\ref{eq:bayes-theorem}.
The probability density function of an observed quantity $d_{i}$
with $1\sigma$ uncertainties $+\sigma_{p,i}$ and $-\sigma_{m,i}$
is then,
\begin{equation}
p(d_{i}|\vec{m})\equiv L_{i}=\frac{2}{\sqrt{2\pi}(\sigma_{m,i}+\sigma_{p,i})}\exp\left[-\frac{\left(d_{i}-d_{i}(\vec{m})\right)^{2}}{2\sigma_{i}^{2}}\right],\label{eq:pdf-gaussian}
\end{equation}
with
\begin{equation}
\sigma_{i}=\begin{cases}
\sigma_{m,i}, & d_{i}\leq d_{i}(\vec{m}),\\
\sigma_{p,i}, & d_{i}>d_{i}(\vec{m}).
\end{cases}\label{eq:sigma}
\end{equation}
If $\sigma_{m,i}=\sigma_{p,i}$, the likelihood function (Eq.~\ref{eq:pdf-gaussian})
reduces to the usual Gaussian distribution function. We assume that
all $n_{\mathrm{obs}}$ observables are statistically independent,
i.e. they are assumed to be uncorrelated.
The full likelihood function $p(\vec{d}|\vec{m})$ needed in Eq.~\eqref{eq:our-bayes-theorem}
is then the product of the individual probability density functions
$L_{i}$ of the observed data $d_{i}$ given the model parameters
$\vec{m}$,
\begin{equation}
p(\vec{d}|\vec{m})=\prod_{i=1}^{n_{\mathrm{obs}}}L_{i}\,.\label{eq:likelihood}
\end{equation}
We further define the usual $\chi^{2}$,
\begin{equation}
\chi^{2}=\sum_{i=1}^{n_{\mathrm{obs}}}\frac{\left(d_{i}-d_{i}(\vec{m})\right)^{2}}{\sigma_{i}^{2}},\label{eq:chi2}
\end{equation}
which is useful later on.

Whenever only lower or upper limits of observables are known, we use
a step function as the likelihood instead of the Gaussian function
from Eq.~\ref{eq:pdf-gaussian}. This means that all values above
the lower limit and all values below the upper limit, respectively,
are equally probable.

In Eq.~\ref{eq:likelihood} we assume that the observables are not correlated, 
which may lead to a loss of information.
For example, effective temperatures and surface gravities that
are inferred from models of stellar spectra calculated with
a stellar atmosphere code are typically correlated. They are correlated
in the sense that fitting a spectral line equally well with a hotter effective
temperature requires a larger gravity. Such correlations are not accounted
for in our current approach because the needed information is usually
not published. The correlations are valuable because they contain
information to constrain stellar parameters better and may thus result
in smaller uncertainties. In principle correlations can readily be 
incorporated in our approach by including the covariance matrix in the likelihood 
function.

\subsection{Prior functions}\label{sec:priors}

The prior functions contain our a priori knowledge of the model parameters
$\vec{m}$ (i.e. $M_{\mathrm{ini}},\, v_{\mathrm{ini}}\,,Z,\,\tau$).
The prior functions do not include our knowledge of stellar evolution
such as how much time stars spend in different evolutionary phases.
This knowledge, that essentially is also a priori, enters our approach
through the stellar models (Sec.~\ref{sec:stellar-model-grids}).
As for the likelihood function, we assume that the individual model
parameters are independent such that the prior function splits into
a product of four individual priors for the four model parameters,
\begin{equation}
p(\vec{m})=\xi(M_{\mathrm{ini}})\,\theta(v_{\mathrm{ini}})\,\psi(Z)\,\zeta(\tau).\label{eq:prior-without-incl}
\end{equation}

Our a priori knowledge of initial masses, i.e. the initial mass prior
$\xi(M_{\mathrm{ini}})$, is given by the initial mass function. The
initial mass function is commonly expressed as a power-law function,
\begin{equation}
\xi(M_{\mathrm{ini}})\propto M_{\mathrm{ini}}^{\gamma},\label{eq:imf}
\end{equation}
with slope $\gamma$. The slope is $\gamma=-2.35$ for a Salpeter
initial mass function \citep{1955ApJ...121..161S}, meaning that lower
initial masses are more probable/frequent than higher. Alternative
parameterisations of the initial mass function may be used as well
\citep[see e.g. the review by][]{2010ARA&A..48..339B}. A uniform
mass function, i.e. $\xi(M_{\mathrm{ini}})=\mathrm{const.}$, may
be used if no initial mass shall be preferred a priorly. 

As an initial rotational velocity prior, $\theta(v_{\mathrm{ini}})$,
we use observed distributions of stellar rotational velocities such
as those found by \citet{2008A&A...479..541H} for O and B-type stars
in the Milky Way and Magellanic Clouds. For MW and LMC stars, the
observed \citet{2008A&A...479..541H} distributions follow a Gaussian
distribution with mean of $100\,\mathrm{km}\,\mathrm{s^{-1}}$ and
standard deviation $106\,\mathrm{km}\,\mathrm{s^{-1}}$ and for SMC
stars a Gaussian distribution with mean $175\,\mathrm{km}\,\mathrm{s^{-1}}$
and standard deviation $106\,\mathrm{km}\,\mathrm{s^{-1}}$. Other
observed rotational velocity distributions of OB stars are equally
well suited as priors; e.g. the distributions found by \citet{1977ApJ...213..438C},
\citet{1997MNRAS.284..265H}, \citet{2002ApJ...573..359A}, \citet{2006A&A...452..273M,2007A&A...462..683M},
\citet{2009ApJ...700..844P}, \citet{2010ApJ...722..605H}, \citet{2013A&A...550A.109D}
or \citet{2013A&A...560A..29R}. We note that the observed distributions
of rotational velocities are not necessarily the initial distributions
that are actually required as prior function \citep[see][]{2013ApJ...764..166D}.
A uniform prior, i.e. every initial rotational velocity is a priori
equally probable, may be used as well. 

The metallicity prior $\psi(Z)$ has yet no influence because the
metallicity is not a free model parameter in the currently supported
stellar model grids. In principle, any probability distribution such
as a uniform or Gaussian distribution is suited to describe a priori
knowledge of the metallicity of a star under investigation. The prior
can, for example, encompass knowledge of a spread in metallicity of
a population of stars \citep[e.g.][]{2014A&A...565A..89B}.

The age prior $\zeta(\tau)$ is set by the star formation history
of the region to which an observed star belongs. If the observed star
is a member of a star cluster that formed in a single starburst, the
age prior may be a Gaussian distribution with mean equal to the age
of the cluster and width corresponding to the duration of the star
formation process. A uniform age prior corresponds to assuming a constant
star formation rate in the past such that all ages are a priori equally
probable.

It is typically assumed that the inclination of a star with respect
to our line-of-sight does not affect observables except for projected
rotational velocities, $v\sin i$. This assumption breaks down for
rapid rotators because their equators are cooler than their poles
as a result of a distortion of the otherwise spherically symmetric shape
of a star by the centrifugal force. The inferred effective temperature
and luminosity are then a function of the inclination angle of the
star towards our line-of-sight The latter effect is beyond the scope
of this paper. However, projected rotational velocities are often
determined from stellar spectra. Whenever $v\sin i$ measurements
are compared to stellar models, we take the equatorial rotational
velocities of the models and combine them with any possible orientation
of the star in space, i.e. with any possible inclination angle, to
derive the $v\sin i$ values. By doing so, we introduce a fifth model
parameter, the inclination angle $i$, and have to define the appropriate
inclination prior, $\phi(i)$. Our assumption that the rotation axes
of stars are randomly oriented in space results in inclination angles
that are not equally probable. It is, for example, more likely that
a star is seen equator-on ($i=90^{\circ}$) than pole-on ($i=0^{\circ}$)
because the solid angle of a unit sphere, $\mathrm{d}\Omega\propto\sin i\,\mathrm{d}i$,
of a region around the equator is larger than that of a region around
the pole. The inclination prior is then,
\begin{equation}
\phi(i)=\sin i,\label{eq:inclination-prior}
\end{equation}
and the full prior in Eq.~\eqref{eq:our-bayes-theorem} reads 
\begin{equation}
p(\vec{m})=\xi(M_{\mathrm{ini}})\,\theta(v_{\mathrm{ini}})\,\psi(Z)\,\zeta(\tau)\,\phi(i).\label{eq:prior-with-incl}
\end{equation}

Note that the functional form of the priors may predict non-zero probabilities
for values of the model parameters that are physically meaningless.
A Gaussian $v_{\mathrm{ini}}$ prior may result in non-zero probabilities
for negative rotational velocities and a Gaussian age prior in non-zero
probabilities for negative ages. Ideally, the prior functions should
be properly truncated and renormalised to allow only for meaningful
values of the model parameters. However, we can skip this step because
the stellar model grids ensure that we only analyse physically meaningful
values of the model parameters and the application of Bayes' theorem
as in Eq.~\eqref{eq:our-bayes-theorem} requires a proper renormalisation
anyhow such that we can simply work with priors that are not properly
normalised but describe relative differences correctly.

By default, \bonnsai assumes a Salpeter initial mass prior, uniform priors
in $v_\mathrm{ini}$, $Z$ and age and that stellar rotation axes are randomly
oriented in space.

\subsection{Stellar model grids}\label{sec:stellar-model-grids}

Currently, \bonnsai supports the stellar models of \citet{2011A&A...530A.115B}
and \citet{Koehler+2014} summarised in Table~\ref{tab:mass-and-velocity-ranges}.
\bonnsai follows a grid-based approach, i.e. we compute the posterior
probability distribution by sampling models from a dense, precomputed
and interpolated grid. The model grids have a resolution in initial
mass of $\Delta M_{\mathrm{ini}}=0.2\,\msun$, in age of $\Delta\tau=0.02\,\mathrm{Myr}$
and in initial rotational velocity of $\Delta v_{\mathrm{ini}}=10\,\mathrm{km}\,\mathrm{s}^{-1}$.
The model grids Bonn MW, LMC and SMC with their different initial
mass coverage contain about 7.5, 20 and 8 million stellar models,
respectively. If projected rotational velocities are matched to stellar
models, we probe ten inclination angles from $0^{\circ}$ to $90^{\circ}$.
The model grids are computed by a linear interpolation of the stellar
models of \citet{2011A&A...530A.115B} and \citet{Koehler+2014} with
the technique described in \citet{2011A&A...530A.116B}.

The stellar model grids contain our (a priori) knowledge of stellar
evolution. For example, the density of the stellar models is indicative
of the time spent by the models in different evolutionary phases and
thus ensures that this knowledge is properly taken into account in
our approach.

\subsection{Goodness-of-fit}\label{sec:goodness-of-fit}

\begin{figure*}
\begin{centering}
\includegraphics[width=6cm]{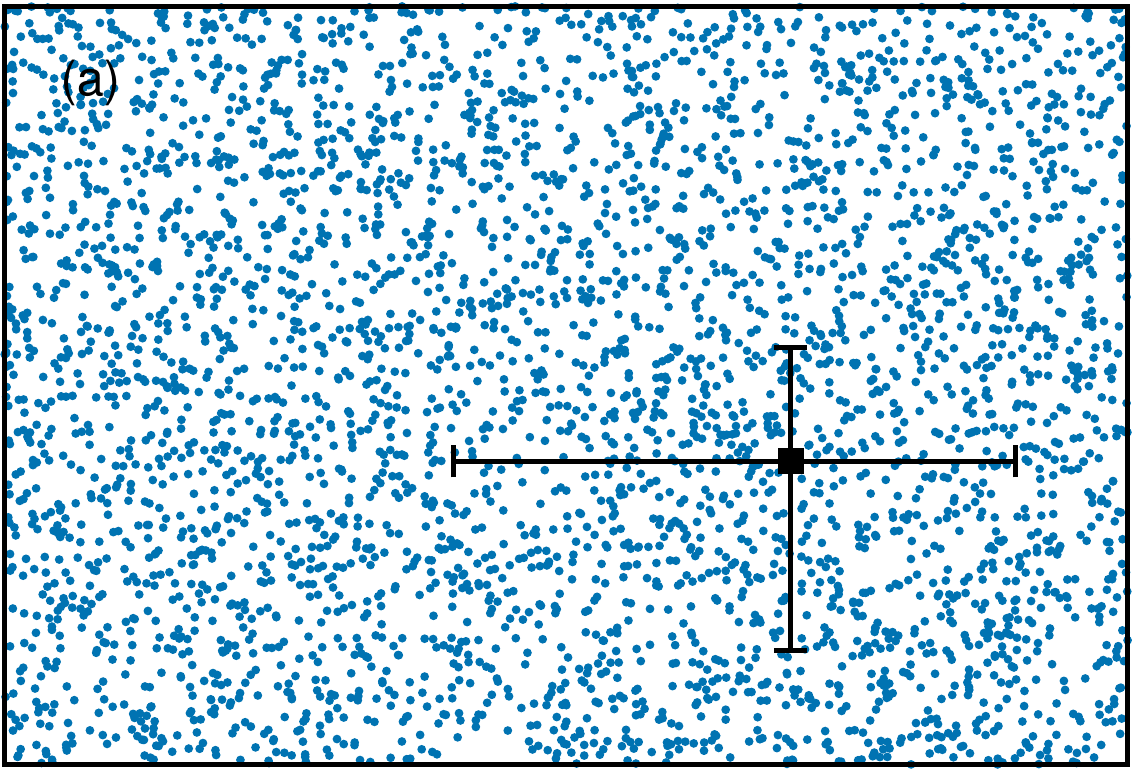}\hspace{0.2cm}\includegraphics[width=6cm]{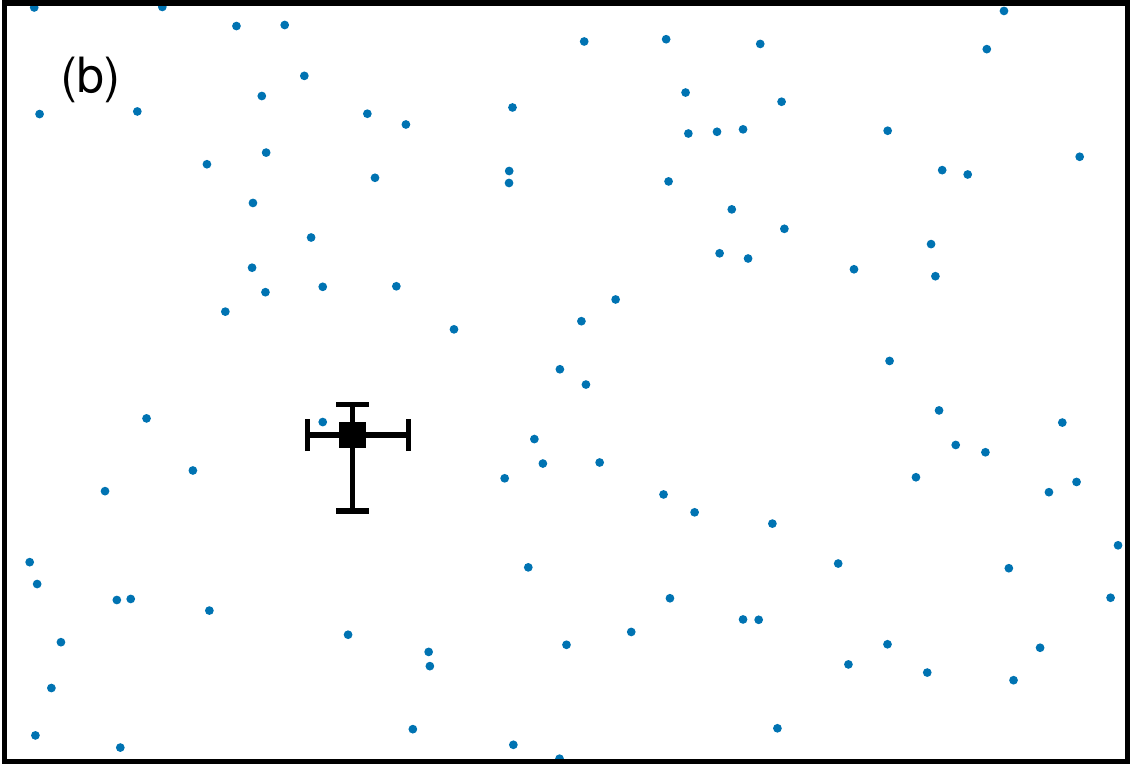}\hspace{0.2cm}\includegraphics[width=6cm]{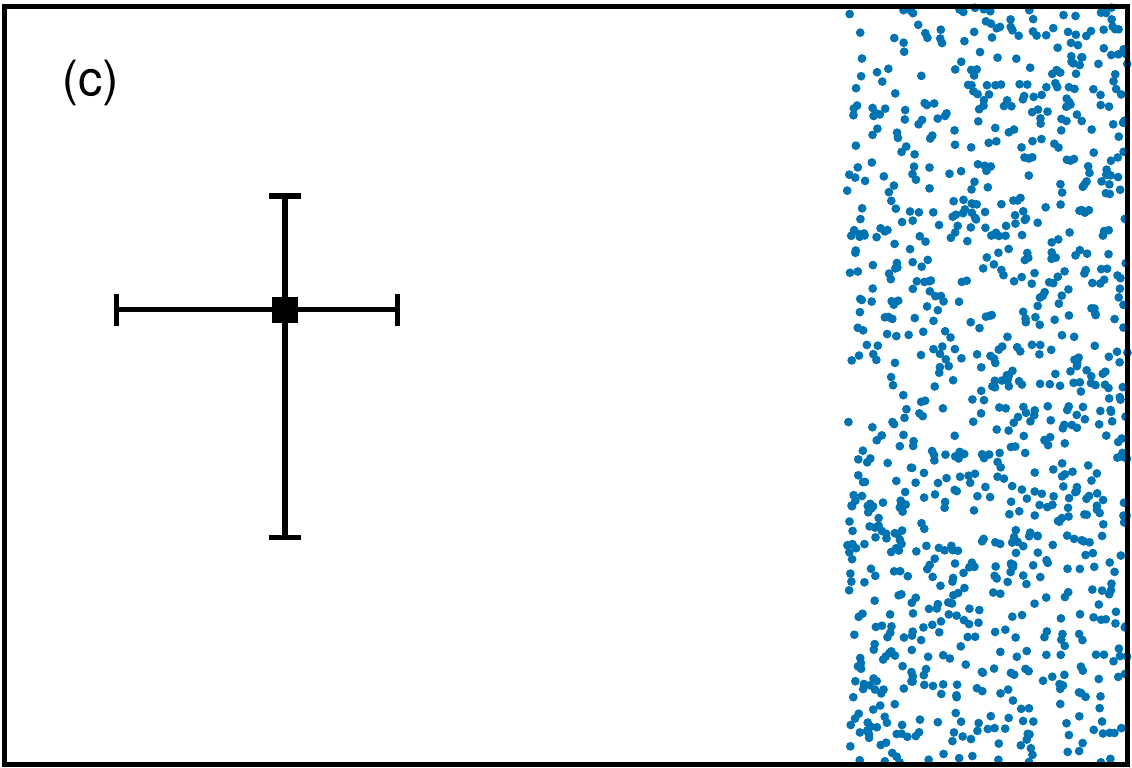}
\par\end{centering}

\caption{Hypothetical model grid coverage of three observations. The resolution
of the model grid is high compared to the observed uncertainties in
the left panel (a) and the models sample the observation well. The
model grid is sparse in the middle panel (b) such that the model density
is not enough to resolve the $1\sigma$ uncertainties of the observations
--- reliable model parameters cannot be determined. In the right panel
(c), the models are physically unable to match the observations, i.e.
the resolution of the model grid is high compared to the observed
uncertainties but the models do not probe the observed region of the parameter
space.}

\label{fig:demo-model-resolution}
\end{figure*}

A crucial aspect of any fitting procedure is to check the goodness
of the fit (e.g.\ when fitting a straight line to data points that 
may actually follow an exponential distribution). 
Typically the $\chi^{2}$ of the fit is used as the goodness of the fit. Similarly,
we have to check whether the stellar models fitted to observations
are a good representation of the observations for the determined model
parameters $\vec{m}$. 

As an example, consider a star in the HR diagram that lies more than
$5\sigma$ bluewards of the zero-age main-sequence and is compared
to non-rotating, main-sequence single star models. Our Bayesian approach
presented so far would return model parameters although the models
are actually unable to reach the observed position of the star in
the HR diagram. 

In general, the stellar models may fail to match an observation because
\begin{itemize}
\item the star is not covered by the underlying stellar model grid,
\item physics is missing in the stellar models, e.g. magnetic fields, a
different treatment of rotation and rotational mixing or duplicity,
\item there are problems with the calibrations of e.g. convective core overshooting,
the efficiency of rotational mixing or stellar winds,
\item there are difficulties with the methods from which observables like
the surface gravity are derived (e.g. stellar atmosphere models),
\item there are problematic observations like unseen binary companions that
lead to misinterpretation of fluxes and spectra.
\end{itemize}
The goodness of a fit is often checked by eye. Besides a by-eye method,
we develop objective and robust tests that allow us to reject the
determined model parameters for a given significance level. By default,
we use a significance level of $\alpha=5\%$ in our tests as commonly
adopted in statistics.

Within our approach, stellar models might be unable to match observations
not only because of the reasons given above but also because of a
too low resolution of the model grid. We illustrate these cases in
Fig.~\ref{fig:demo-model-resolution}. The left panel of Fig.~\ref{fig:demo-model-resolution}
contains a hypothetical observation including error bars and a hypothetical
coverage of the observed parameter space by a stellar model grid.
It is evident that the models densely cover the observation. The resolution
of the model grid, i.e. the average distance between adjacent models,
is small compared to the observational uncertainties. In the middle
panel of Fig.~\ref{fig:demo-model-resolution}, the stellar model
grid in principle covers the observations but the model grid is sparse
compared to the observational uncertainties. In order to determine
reliable model parameters, the resolution of the model grid needs
to be increased. Such situations are encountered whenever the observational
uncertainties are so accurate that the observations including their
error bars fall in between model grid points. In the right panel of
Fig.~\ref{fig:demo-model-resolution}, the model grid is dense compared
to the observational uncertainties but the models are unable to match
the observations. The resolution of the model grid could be infinite
but the models would still not reproduce the observations. We not
only want to identify situations as in Figs.~\ref{fig:demo-model-resolution}b
and~c but also want to be able to distinguish between these situations.

To test whether the resolution of the stellar model grid is sufficient
to determine reliable model parameters, we investigate the average
spacing of those ten models that are closest to the best-fitting model.
We take the nearest neighbours of each of the eleven models and compute
the average differences of each observable and the average $\chi^{2}$
of these pairs. We define as our resolution criterion that the average
differences are less than one fifth of the corresponding $1\sigma$
uncertainties. This ensures that there are about ten models within
$1\sigma$ of each observable and that we know how significant the
$\chi^{2}$ of the best fitting model is with respect to the resolution
of the stellar model grid. 

Once it is ensured that the resolution of the model grid is sufficient,
we test whether the stellar models are able to match the observations.
One straightforward way to judge the goodness of the estimated parameters
is to do a classical $\chi^{2}$-hypothesis test (Pearson's $\chi^{2}$-hypothesis
test). The $\chi^{2}$ of the best-fitting model, $\chi_{\mathrm{best}}^{2}$,
is compared to the maximum allowable $\chi_{\mathrm{max}}^{2}$ for
a given significance level $\alpha$, where the maximum allowable
$\chi_{\mathrm{max}}^{2}$ is such that the integrated probability
of the $\chi^{2}$-distribution for $\chi^{2}\geq\chi_{\mathrm{max}}^{2}$
is equal to the significance level. This means that there is a probability
less than $\alpha$ that $\chi_{\mathrm{best}}^{2}\geq\chi_{\mathrm{max}}^{2}$
for the best-fitting stellar model. The $\chi^{2}$-distribution is
defined by the degrees of freedom which, in our case, is given by
the number of observables.
If some observables are dependent on each other, i.e.\ if some observables can be derived from others
as is the case for surface gravity, mass and radius or luminosity, effective temperature and radius, 
the $\chi^{2}$-test gives a too large $\chi_{\mathrm{max}}^{2}$, i.e. the test is performed with an effectively
smaller significance level.

The $\chi^{2}$-test only incorporates the best-fitting stellar model.
In a Bayesian analysis there are not only the model parameters of
the best-fitting stellar model but the full posterior probability
distribution of the model parameters that do not necessarily peak
at the best-fitting model parameters. We make use of this by conducting
a posterior predictive check. The idea is to compare the predictions
of the stellar models for the estimated model parameters to the observations
to check whether the predictions are in agreement with the observations.
If they are not in agreement, the estimated model parameters and hence
the stellar models cannot reproduce/replicate the observations. In
Bayesian statistics, the predictions of the models for the observables
are called replicated observables, $\vec{d}_{\mathrm{rep}}$, and
are computed from the full posterior probability distribution, $p(\vec{m}|\vec{d})$,
\begin{equation}
p(\vec{d}_{\mathrm{rep}}|\vec{d})=\int_{\vec{m}}\, p(\vec{d}_{\mathrm{rep}}|\vec{m})p(\vec{m}|\vec{d})\,\mathrm{d}\vec{m},\label{eq:replicated-model-parameters}
\end{equation}
where $p(\vec{d}_{\mathrm{rep}}|\vec{d}$) is the probability distribution
of the replicated observables (given the original observables) and
$p(\vec{d}_{\mathrm{rep}}|\vec{m})$ is the probability of the replicated
parameters $\vec{d}_{\mathrm{rep}}$ given a stellar model with parameters
$\vec{m}$ (i.e. $p(\vec{d}_{\mathrm{rep}}|\vec{m})$ consists, in
our case, of delta-functions). From the likelihood function (Eq.~\ref{eq:likelihood})
and the posterior predictive probability distribution of the replicated
observables (Eq.~\ref{eq:replicated-model-parameters}), we compute
the probability distributions $p(\Delta d_{i}|\vec{d})$ of the differences
between the replicated and original observables, $\Delta d_{i}\equiv d_{\mathrm{rep},i}-d_{i}$.
If the stellar models can reproduce the observations, the differences
between replicated and original observables have to be consistent
with being zero. We define the differences to be consistent with zero
if the integrated probabilities for $\Delta d_{i}>0$ and $\Delta d_{i}<0$
are both larger than the significance level $\alpha$ for all observables
$d_{i}$, $i=1\dots n_{\mathrm{obs}}$, i.e.
\begin{eqnarray}
\int_{\Delta d_{i}<0}\, p(\Delta d_{i}|\vec{d})\,\mathrm{d}d_{i} & \geq & \alpha\;\land\;\int_{\Delta d_{i}>0}\, p(\Delta d_{i}|\vec{d})\,\mathrm{d}d_{i}\geq\alpha,\;\forall i\,.\label{eq:our-bayes-gof-definition}
\end{eqnarray}
In other words, we say that the stellar models cannot reproduce the
observations if the probability that the model prediction of any quantity
is larger or smaller than the observational value is $\geq95\%$ ($\geq1-\alpha$
with $\alpha=5\%$). 

Besides these automated and objective tests, we check the coverage
of the observed parameter space graphically in diagrams similar to
the illustrations in Fig.~\ref{fig:demo-model-resolution}. We place
one dot for each stellar model together with the observables and their
uncertainties into two-dimensional projections of the parameter space
of the observables. This results in $n_{\mathrm{obs}}(n_{\mathrm{obs}}+1)/2$
projections for $n_{\mathrm{obs}}$ observables from which the resolution
can be assessed as well as whether the models are able to reproduce
the observations (see Fig.~\ref{fig:non-existent-star-parameter-space-coverage}
below for an example).

\subsection{Our new approach in practice}\label{sec:bonnsai-in-practice}

In practice we perform the following steps:
\begin{enumerate}
\item We select all stellar models from a database (model grid) that are
within $5\sigma_{i}$ of all observables. The pre-selection reduces
the parameter space of the stellar models $\left(M_{\mathrm{ini}},v_{\mathrm{ini}},Z,\tau,i\right)$
that needs to be scanned and thus accelerates the analysis.
\item We scan the pre-selected model parameter space and compute for each
model the posterior probability according to Eq.~\eqref{eq:our-bayes-theorem}.
The posterior probability consists of the likelihood from Eq.~\eqref{eq:likelihood}
and a weighting factor that gives the probability of finding a stellar
model with the given model parameters, $\vec{m}$. The weighting factor
consists of two contributions: first, it takes into account the prior
function from Eq.~\ref{eq:prior-with-incl} factoring in our a priori
knowledge about the probability of finding a particular stellar model
and, second, the volume $\Delta V=\Delta M_{\mathrm{ini}}\,\Delta v_{\mathrm{ini}}\,\Delta Z\,\Delta\tau\,\Delta i$
that a particular stellar model covers in the model grid. The latter
allows us to use non-equidistant model grids. 
\item We renormalise the posterior probabilities such that Eq.~\eqref{eq:normalisation}
is fulfilled and create 1D and 2D probability functions/maps for the
model parameters $\vec{m}$ by marginalisation, i.e. by projecting
the posterior probability distribution onto, e.g., the $M_{\mathrm{ini}}$
axis or into the $v_{\mathrm{ini}}-M_{\mathrm{ini}}$ plane. Additionally
we use Eq.~\eqref{eq:replicated-model-parameters} to derive probability
functions/maps of any stellar parameter to e.g. predict yet unobserved
surface nitrogen abundances of the stellar models for the estimated
model parameters.
\item The resolution and goodness-of-fit tests are conducted.
\item The 1D probability functions are analysed to compute the mean, median
and mode including their $1\sigma$ uncertainties.
\end{enumerate}

\section{Testing \bonnsai with mock stars}\label{sec:mock-stars}

\begin{figure}
\begin{centering}
\includegraphics[width=9cm]{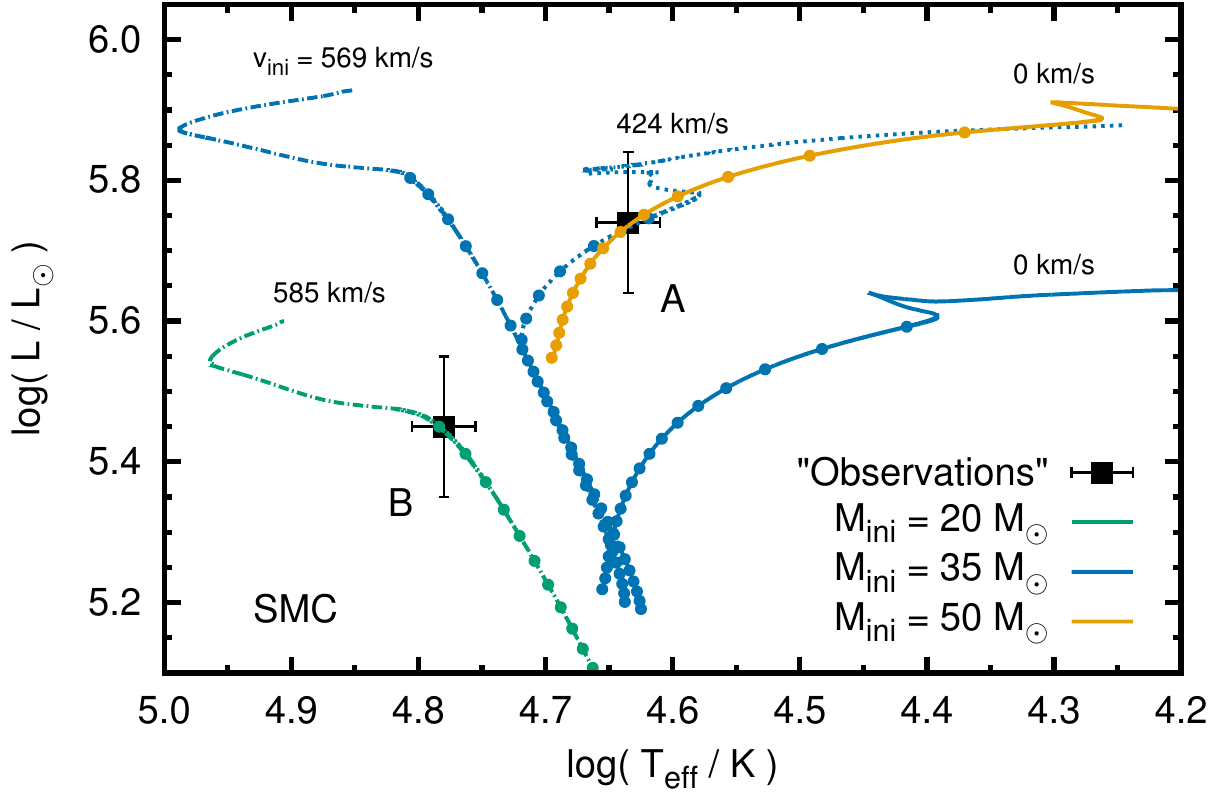}
\par\end{centering}

\caption{Position of the mock stars, Star~A and Star~B, in the HR diagram compared
to rotating and non-rotating stellar evolutionary models of \citet{2011A&A...530A.115B}
of SMC composition. The dots on the stellar tracks are equally spaced
by $0.25\,\mathrm{Myr}$.}

\label{fig:mock-stars-hrd}
\end{figure}

In the following sections, we apply \bonnsai to the two SMC mock stars,
Star~A and Star~B, whose position in the HR diagram is shown in
Fig.~\ref{fig:mock-stars-hrd} together with SMC stellar evolutionary
models of \citet{2011A&A...530A.115B}. We analyse Star A in Sec.~\ref{sec:mock-star-a}
and Star B in \ref{sec:mock-star-b}.

\subsection{Mock Star A}\label{sec:mock-star-a}

Mock Star A has an effective temperature of $T_{\mathrm{eff}}=43200\pm2500\,\mathrm{K}$
and a luminosity of $\log L/\lsun=5.74\pm0.10$ (Fig.~\ref{fig:mock-stars-hrd}).
Its position in the HR diagram is equally well matched by an initially
rapidly ($424\,\mathrm{km}\,\mathrm{s^{-1}}$) rotating $35\,\msun$
star and a non-rotating $50\,\msun$ star. The initial rapid rotator
starts out evolving chemically homogeneously. Rotationally induced
mixing brings helium, synthesized by hydrogen burning in the stellar
core, to the surface which in turn reduces the electron scattering
opacity. Therefore, it stays more compact, evolves towards hotter
effective temperatures and becomes more luminous than its non-rotating
counterpart. It is tempting to conclude that both evolutionary scenarios
are equally likely because both fit the position of the star in the
HR diagram equally well. However, this is a biased view that does
not take a priori knowledge of stars and stellar evolution into account
--- we can actually exclude the rapidly rotating $35\,\msun$ star
with more than 95\% confidence as we show below. 

There are two different sources of a priori knowledge, namely that
of (a) the model parameters, i.e. what we know about the initial mass,
age, initial rotational velocity and metallicity of the star before
analysing it, and (b) stellar evolution. The former enters our approach
through the prior functions (Sec.~\ref{sec:priors}) and the latter
through the stellar models used to compute the likelihood (Sec.~\ref{sec:likelihood}).
From the point of view of the prior functions, $35\,\msun$ models
are preferred over $50\,\msun$ because of the IMF, but moderately
rotating $50\,\msun$ models over rapidly rotating $35\,\msun$ stars
because of observed distributions of stellar rotation rates. From
the point of view of stellar models, the $50\,\msun$ track is preferred
because the $50\,\msun$ model spends more time at the observed position
in the HR diagram than the $35\,\msun$ model that is close to the
end of its main-sequence evolution where evolution is more rapid.
The $50\,\msun$ models are also preferred because there are many
models of that mass with different initial rotational velocities that
reach the observed position in the HR diagram while there is only
a narrow range of initial rotation rates of $35\,\msun$ models that
match the observations.

The a priori knowledge allows us to quantify how likely both models
are. Only additional observables that are sensitive to rotation enable
us to fully resolve the degenerate situation.

In the following, we match Star A against the SMC models of \citet{2011A&A...530A.115B},
choose a Salpeter mass function as initial mass prior and a uniform
age prior. We vary the initial rotational velocity prior and use an
additional constraint on the surface helium abundance to show their
influence on the posterior probability distributions. The resolution
test, the $\chi^{2}$-hypothesis test and the posterior predictive
checks are passed in all cases. We present a summary of our test cases
in Table~\ref{tab:basic-examples-summary}.

\subsubsection{Uniform initial rotational velocity prior}\label{sec:basic-example-uniform-vini-prior}

\begin{figure}
\begin{centering}
\includegraphics[width=9.0cm]{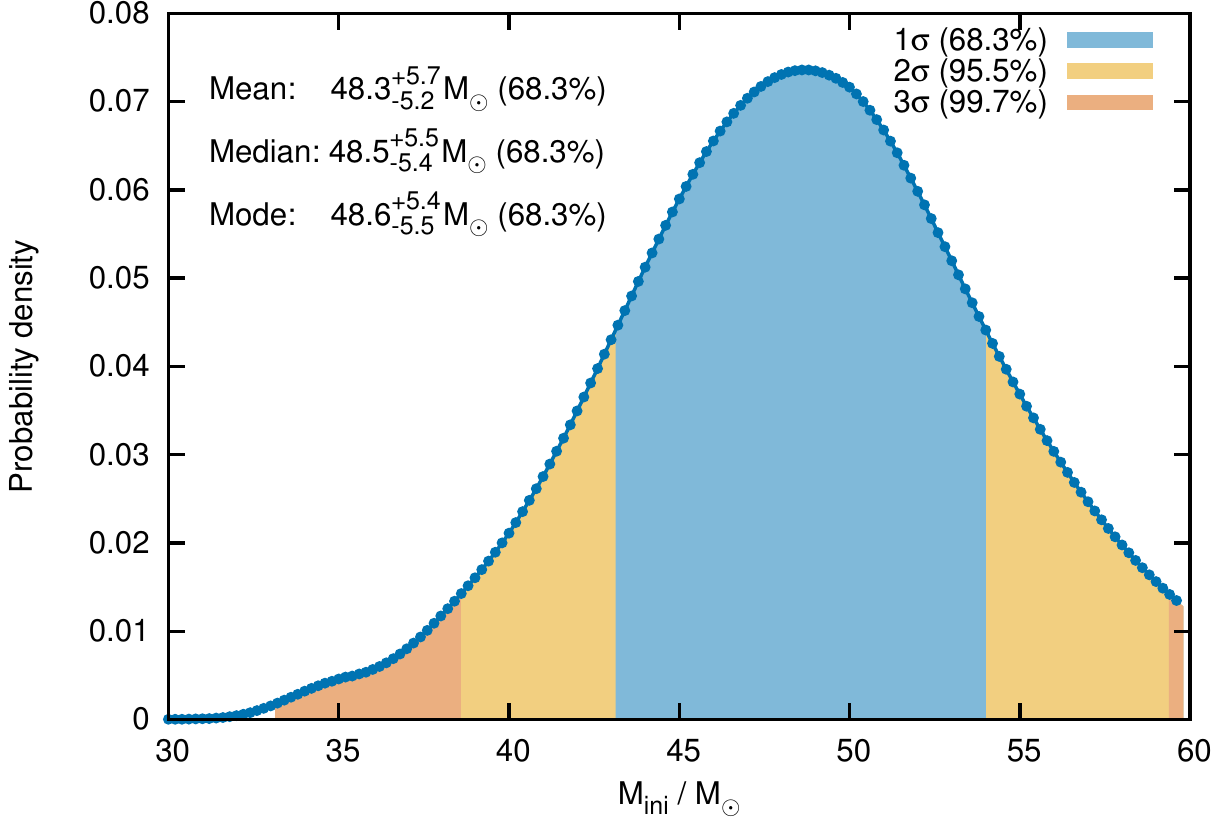}
\par\end{centering}

\caption{Initial mass posterior probability distribution of mock Star A from
Fig.~\ref{fig:mock-stars-hrd}. The shaded regions give the $1\sigma$,
$2\sigma$ and $3\sigma$ confidence regions. }

\label{fig:basic-example-mini}
\end{figure}

\begin{figure}
\begin{centering}
\includegraphics[width=9cm]{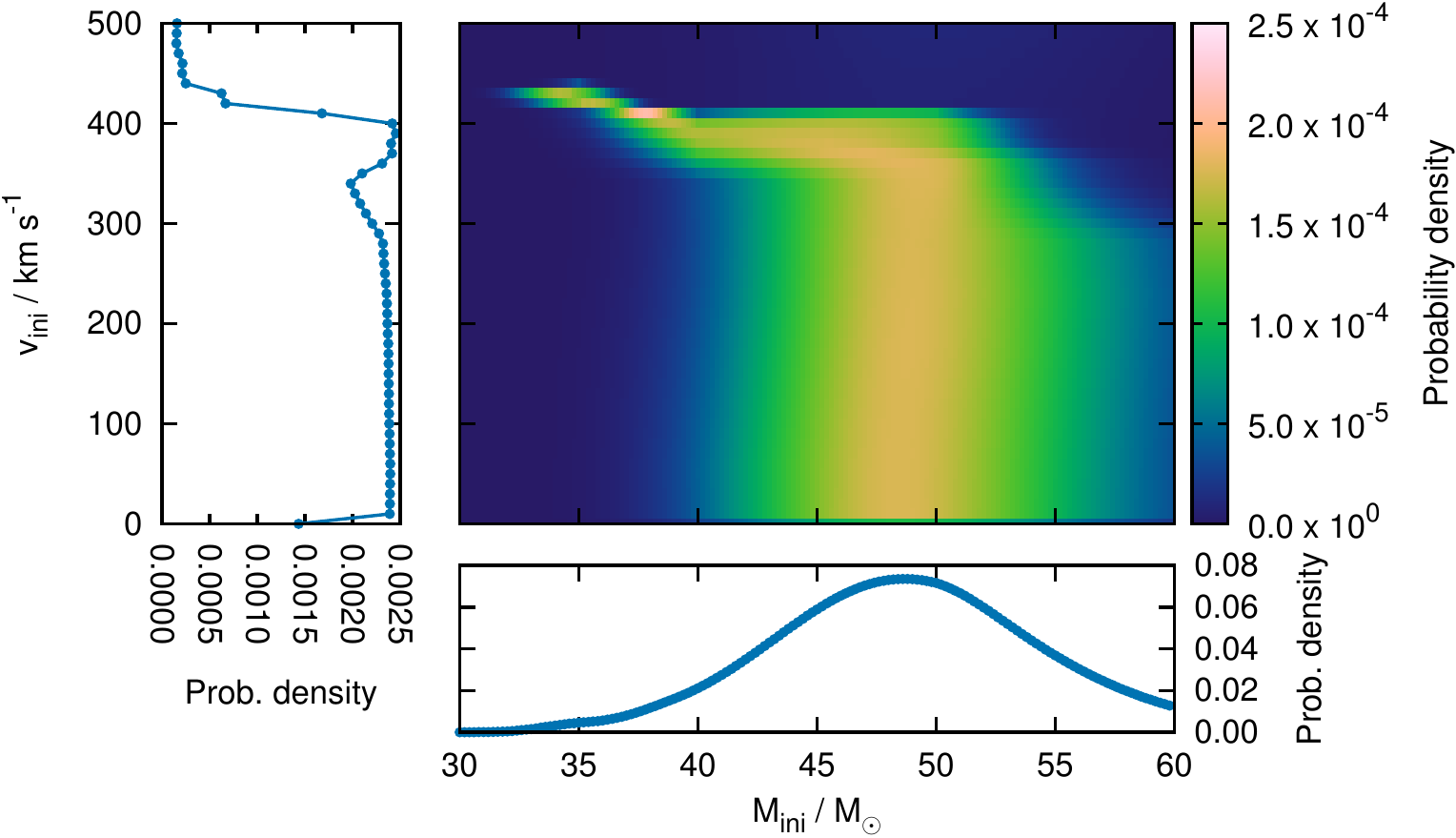}
\par\end{centering}

\caption{Posterior probability map of the $v_{\mathrm{ini}}-M_{\mathrm{ini}}$
plane (middle panel) adopting a Salpeter initial mass function as
$M_{\mathrm{ini}}$ prior and a uniform prior for $v_{\mathrm{ini}}$.
One dimensional posterior probability distributions of initial rotational
velocities $v_{\mathrm{ini}}$ (left panel) and initial masses $M_{\mathrm{ini}}$
(bottom panel) are also given.}

\label{fig:basic-example-vini-mini}
\end{figure}

At first we apply a uniform $v_{\mathrm{ini}}$ prior, i.e. we assume
that all initial rotational velocities are a priori equally probable.
The resulting posterior probability distribution of the initial mass
is shown in Fig.~\ref{fig:basic-example-mini} as a histogram with
a binwidth of $0.2\,\msun$. The mean, median and mode are given with
their $1\sigma$ uncertainties (the confidence level is indicated
in parentheses). The most probable initial mass is $48.6_{-5.5}^{+5.4}\,\msun$,
the most probable age is $2.3_{-0.6}^{+0.5}\,\mathrm{Myr}$ and the
initial rotational velocity is unconstrained, i.e. its posterior probability
distribution is uniform until it steeply drops-off at $v_{\mathrm{ini}}\gtrsim400\,\mathrm{km}\,\mathrm{s}^{-1}$
(left panel in Fig.~\ref{fig:basic-example-vini-mini}). The likelihood
of a $35\,\msun$ star is small. It is only within the $3\sigma$
uncertainty despite the Salpeter IMF prior that prefers lower initial
masses. The reason why the $50\,\msun$ models are favoured over rapidly
rotating $35\,\msun$ stars is best seen in the $v_{\mathrm{ini}}-M_{\mathrm{ini}}$
plane of the posterior probability distribution in Fig.~\ref{fig:basic-example-vini-mini}.
Only a small subset of $35\,\msun$ fast rotators, those with $v_{\mathrm{ini}}\gtrsim400\,\mathrm{km}\,\mathrm{s^{-1}}$,
reach the observed position in the HR diagram and thus contribute
to the posterior probability distribution of the initial mass in Fig.~\ref{fig:basic-example-mini}.
The range of initial rotational velocities of stars around $50\,\msun$
that contribute to the posterior probability is much larger and has
thus a correspondingly larger weight (Fig.~\ref{fig:basic-example-vini-mini}).
In conclusion, both the $35$ and $50\,\msun$ models reproduce the
position in the HR diagram equally well but it is much more likely
that the star is a $50\,\msun$ star: $35\,\msun$ models are excluded
with a confidence of more than $97.5\%$.

As evident from this example, the (marginalised) posterior distributions
contain a wealth of information that is partly lost when looking only
at the summary statistics, e.g.\ the mode and confidence levels. We
therefore encourage all \bonnsai users to first inspect the
marginalised posterior probability distributions before making use
of the summary statistics.

\subsubsection{Gaussian initial rotational velocity prior}\label{sec:basic-example-other-vini-prior}

\begin{figure}
\begin{centering}
\includegraphics[width=9cm]{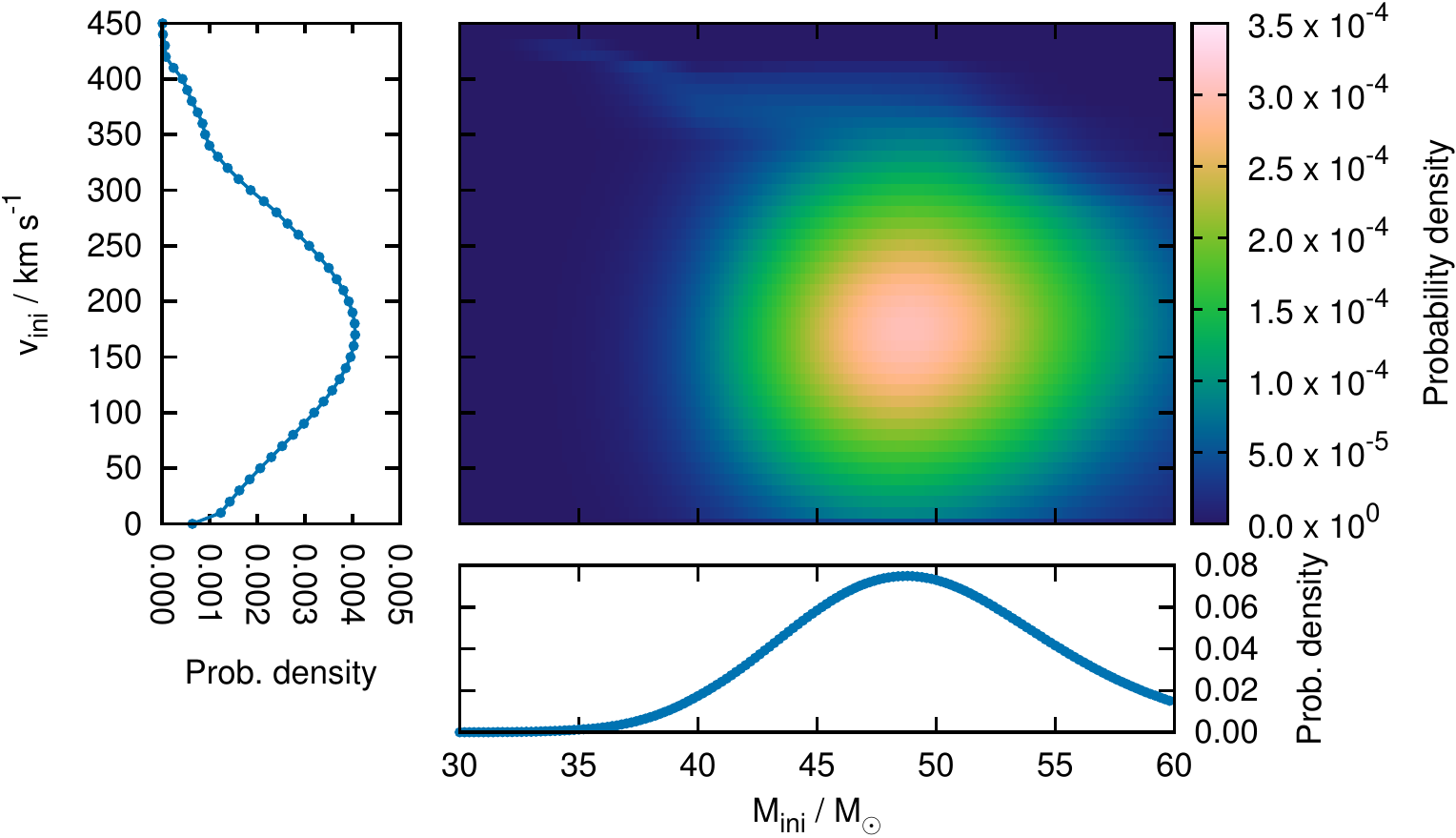}
\par\end{centering}

\caption{As Fig.~\ref{fig:basic-example-vini-mini} but using as $v_{\mathrm{ini}}$
prior the observed distribution of rotational velocities of SMC early
B-type stars from \citet{2008A&A...479..541H}.}

\label{fig:basic-example-other-vini-prior-vini-mini}
\end{figure}

We now change the initial rotational velocity prior to show its influence
on the posterior probability. We use the observationally determined
distribution of rotational velocities of SMC early B-type stars from
\citet{2008A&A...479..541H} as a prior. This distribution is well
approximated by a Gaussian with mean $\left\langle v_{\mathrm{rot}}\right\rangle =175\,\mathrm{km}\,\mathrm{s^{-1}}$
and standard deviation $\sigma_{\mathrm{v}}=106\,\mathrm{km}\,\mathrm{s^{-1}}$.
The distribution disfavours slow ($\lesssim100\,\mathrm{km}\,\mathrm{s^{-1}}$)
and fast rotators ($\gtrsim250\,\mathrm{km}\,\mathrm{s^{-1}}$) compared
to the uniform $v_{\mathrm{ini}}$ prior. This is reflected in the
$v_{\mathrm{ini}}-M_{\mathrm{ini}}$ plane of the posterior probability
in Fig.~\ref{fig:basic-example-other-vini-prior-vini-mini}. Slow
and fast rotators are now less likely than before with the uniform
$v_{\mathrm{ini}}$ prior. The posterior probability function of the
initial mass and age are however nearly unaffected: the most likely
initial mass is $48.8_{-5.1}^{+5.4}\,\msun$. Fast-rotating $35\,\msun$
stars are now even less likely because the chosen prior favours moderate
rotators. The most likely initial rotational velocity now is $170_{-94}^{+99}\,\mathrm{km}\,\mathrm{s}^{-1}$,
i.e. it follows the $v_{\mathrm{ini}}$ prior both in the mean value
and the uncertainty because it is otherwise unconstrained in this
example (Sec.~\ref{sec:basic-example-uniform-vini-prior}).

\subsubsection{Including the surface helium mass fraction}\label{sec:basic-example-breaking-degeneracy}

The degeneracy in HR diagrams (Fig.~\ref{fig:mock-stars-hrd}) that
arises from rotational mixing can be removed if observables that are
sensitive to rotation are incorporated in the analysis. We use observed
surface helium mass fractions, $Y$, as an additional constraint to
demonstrate this (the initial mass fraction of the SMC models of \citet{2011A&A...530A.115B}
is $Y_{\mathrm{ini}}=0.2515$). The surface helium abundance is sensitive
to rotation as rotational mixing brings more helium to the surface
the faster the star rotates. 

A surface helium mass fraction of $Y\leq0.3$ rules out rapid rotators
($\gtrsim350\,\mathrm{km}\,\mathrm{s^{-1}}$) because their surfaces
are enriched in helium to more than 30\% in mass. When assuming a
uniform $v_{\mathrm{ini}}$ prior, the most likely initial mass is
$49.2_{-5.5}^{+4.7}\,\msun$, i.e. slightly larger than before because
rapid rotators are totally excluded and not only suppressed as in
Sec.~\ref{sec:basic-example-other-vini-prior} without the additional
surface helium mass fraction constraint and a Gaussian $v_{\mathrm{ini}}$
prior.

Contrarily, a surface helium mass fraction of $Y\geq0.4$ allows only
for rapid rotators because stars rotating initially slower than about
$350\,\mathrm{km}\,\mathrm{s^{-1}}$ do not enrich their surfaces
by more than 40\% in helium at the observed position in the HR diagram.
So even assuming that the initial rotational velocities are distributed
according to \citet{2008A&A...479..541H} --- a Gaussian distribution
that highly suppresses rapid rotators --- results in a most likely
initial rotational velocity of $400_{-15}^{+14}\,\mathrm{km}\,\mathrm{s^{-1}}$.
Consequently, the most likely initial mass is lower, namely $39.4_{-3.2}^{+5.7}\,\msun$,
and the most likely age is older, namely $3.9_{-0.7}^{+0.8}\,\mathrm{Myr}$.

\begin{table}
\caption{Summary of the assumptions, additional constraints and resulting model
parameters of our mock Star A (Secs.~\ref{sec:basic-example-uniform-vini-prior}--\ref{sec:basic-example-breaking-degeneracy}).}

\begin{centering}
\begin{tabular}{ccccc}
\toprule 
Add. constraint & $v_{\mathrm{ini}}$ prior & $M_{\mathrm{ini}}$ ($\msun$) & $\tau$ (Myr) & $v_{\mathrm{ini}}$ ($\mathrm{km}\,\mathrm{s}^{-1}$)\tabularnewline
\midrule
\midrule 
-- & uniform & $48.6_{-5.5}^{+5.4}$ & $2.3_{-0.6}^{+0.5}$ & unconstr.\tabularnewline
-- & Hunter '08 & $48.8_{-5.1}^{+5.4}$ & $2.3_{-0.5}^{+0.4}$ & $170_{-94}^{+99}$\tabularnewline
$Y\leq0.3$ & uniform & $49.2_{-5.5}^{+4.7}$ & $2.3_{-0.5}^{+0.4}$ & $\leq350$\tabularnewline
$Y\geq0.4$ & Hunter '08 & $39.4_{-3.2}^{+5.7}$ & $3.9_{-0.7}^{+0.8}$ & $400_{-15}^{+14}$\tabularnewline
\bottomrule
\end{tabular}
\par\end{centering}

\label{tab:basic-examples-summary}
\end{table}

\subsection{Mock Star B}\label{sec:mock-star-b}

Next we consider the mock Star B with a luminosity of $\log L/\lsun=5.45\pm0.10$,
effective temperature $T_{\mathrm{eff}}=60200\pm3500\,\mathrm{K}$
and surface helium mass fraction $Y=0.25\pm0.05$ (Fig.~\ref{fig:mock-stars-hrd}).
Only rapid rotators evolving chemically homogeneously reach the position
of Star B in the HR diagram. Main-sequence SMC models of \citet{2011A&A...530A.115B}
at that position in the HR diagram are significantly enriched with
helium at their surface, $Y=0.89_{-0.09}^{+0.05}$ ($Y_{\mathrm{ini}}=0.2515$).
Contrarily, our mock Star B has the initial helium abundance, thus
the stellar models cannot reproduce the star. In the following, we
show how such situations are robustly identified within \bonnsai using
the goodness-of-fit criteria from Sec.~\ref{sec:goodness-of-fit}
after ensuring that the resolution of the model grid is sufficiently
high.

\subsubsection{Resolution and $\chi^{2}$-hypothesis test}\label{sec:star-B-res-chi2-test}

\begin{figure}
\begin{centering}
\includegraphics[width=9cm]{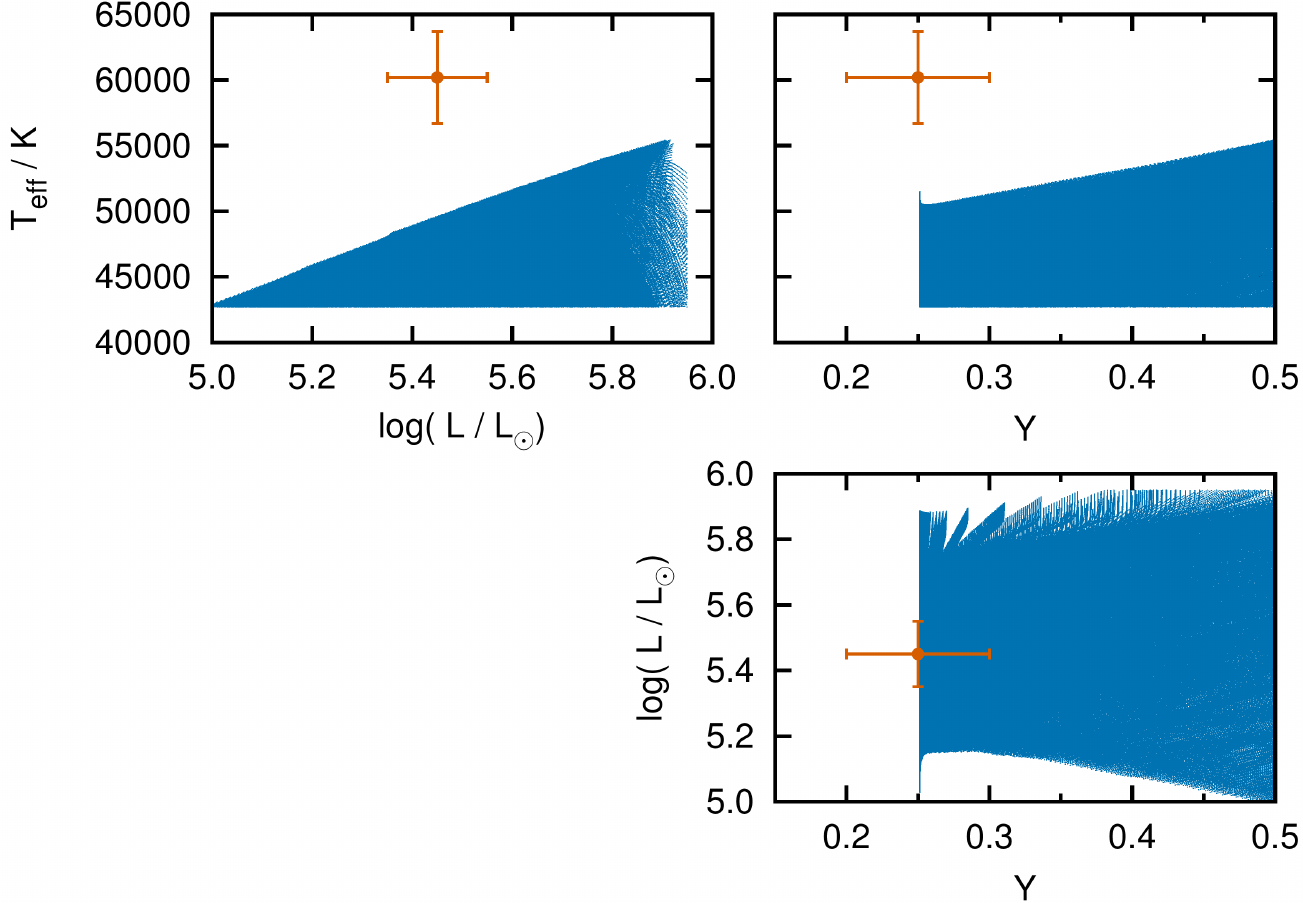}
\par\end{centering}

\caption{Coverage of the projected parameter space of the observables of our
mock Star B (Sec.~\ref{sec:mock-star-b}). One dot for each stellar
model and the observables including their $1\sigma$ uncertainties
are plotted.}

\label{fig:non-existent-star-parameter-space-coverage}
\end{figure}

In Fig.~\ref{fig:non-existent-star-parameter-space-coverage} we
show the stellar model coverage of the parameter space of the observables.
As described in Sec.~\ref{sec:bonnsai-in-practice}, stellar models
are chosen within $5\sigma$ of the observations from the model database,
i.e. the luminosities of the stellar models are in the range $\log L/\lsun=4.95\text{--}5.95$,
the effective temperatures in $T_{\mathrm{eff}}=42700\text{--}77700\,\mathrm{K}$
and the surface helium mass fractions in $Y=0.00\text{--}0.50$. There
are no stellar models in the direct vicinity of the observation because
only those stars that are highly enriched with helium at their surface
($Y=0.89_{-0.09}^{+0.05}$) are found at the observed position in
the HR diagram.

Projections like those in Fig.~\ref{fig:non-existent-star-parameter-space-coverage}
can not be readily used as a criterion to accept a \bonnsai solution
but should only be used as an analysis tool. The best-fitting stellar
model might be far away from the observation (e.g. $5\sigma$) while
all projections homogeneously and densely cover the observation. This
can happen whenever the stellar models cover the surface of the $n_{\mathrm{obs}}$-dimensional
parameter space but not the inside. The projections are then homogeneously
filled with stellar models but the closest model is still far away
from the observation.

In the present case, the automatic resolution test as described in 
Sec.~\ref{sec:goodness-of-fit} confirms the visual impression of 
Fig.~\ref{fig:non-existent-star-parameter-space-coverage} that the 
resolution of the stellar model grid is sufficiently high, i.e.\ that 
the observables, including their error bars, do not fall between model grid points. 
An insufficient resolution as a reason for not finding stellar models close to 
the observation can be excluded and the results of our analysis do not suffer 
from resolution problems.

The maximum allowable $\chi^{2}$
for a significance level $\alpha=5\%$ and three degrees of freedom
is $\chi_{\mathrm{max}}^{2}\approx7.8$. The $\chi^{2}$ of the best-fitting
stellar model is $\chi_{\mathrm{best}}^{2}\approx10.2$. We therefore
conclude that the stellar models are unable to explain the observables with
a confidence of $\geq95\%$.

\subsubsection{Posterior predictive check}

\begin{figure*}
\begin{centering}
\includegraphics[width=18cm]{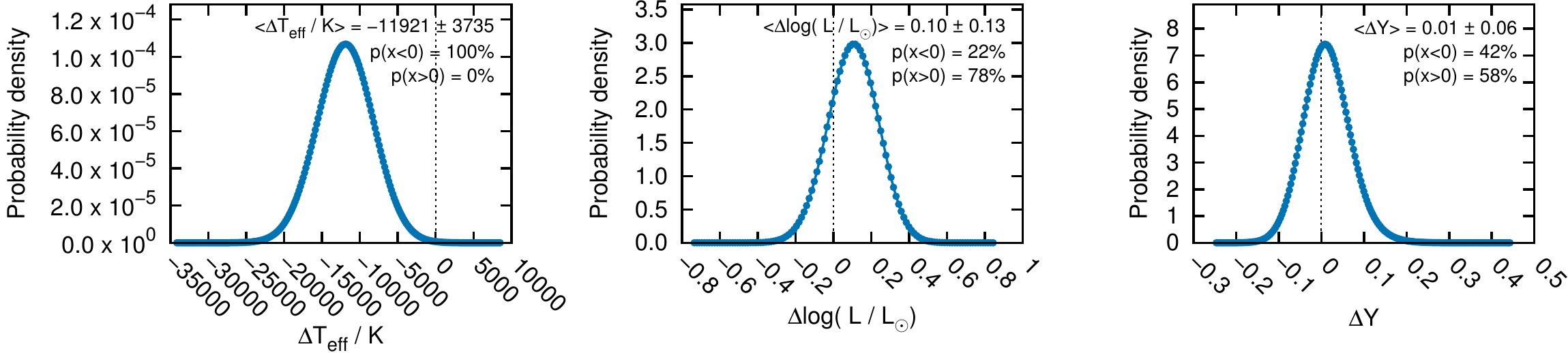}
\par\end{centering}

\caption{Probability distributions of the differences of the replicated and
original observables of our mock Star B (left panel: effective temperature;
middle panel: luminosity; right panel: surface helium mass fraction).
The stellar models predict effective temperatures that are significantly
cooler than the observations and are thus not able to reproduce
the star.}

\label{fig:non-existent-star-posterior-pred-check}
\end{figure*}

In a Bayesian analysis we not only test the best fitting model but
use the full posterior probability distribution to evaluate the goodness
of the fit. From the obtained model parameters, we compute the model
predictions, i.e. the probability distributions, of the observables
effective temperature, luminosity and surface helium mass fraction,
that are called replicated observables. The probability distributions
of the replicated observables are then compared to those of the original
observations, i.e. to the individual likelihood functions, $L_{i}$
(Eq.~\ref{eq:pdf-gaussian}), to check whether the predictions of
the models for the obtained model parameters are in agreement with
the observations. To that end, we compute the probability distributions
of the differences of replicated and original observables which we
show in Fig.~\ref{fig:non-existent-star-posterior-pred-check}. We
find that the effective temperatures deviate by $\Delta T_{\mathrm{eff}}=T_{\mathrm{eff,rep}}-T_{\mathrm{eff,obs}}=-11,920\pm3,735\,\mathrm{K}$
such that the replicated effective temperatures are cooler than the
observed temperatures in more than 99\% of the cases --- the stellar
models can clearly not reproduce the observations. The replicated
luminosities and surface helium mass fractions are in agreement with
the observations ($\Delta\log L/\lsun=0.10\pm0.13$ and $\Delta Y=-0.01\pm0.06$;
Fig.~\ref{fig:non-existent-star-posterior-pred-check}).

\section{Testing stellar evolution models with eclipsing binaries}\label{sec:real-data}

In the previous sections we show that our new method provides stellar
parameters including robust uncertainties and reliably identifies
stars that cannot be reproduced by stellar models when applied to
mock data (Sec.~\ref{sec:mock-stars}). One of the primary goals
of \bonnsai is to test the physics in stellar models by matching the
models to observations in a statistically sound way. For that it is
necessary to have well determined stellar parameters that --- ideally
--- do not rely on extensive modelling and/or calibrations. Eclipsing,
double-lined spectroscopic binaries and interferometric observations
of single stars are prime candidates for this purpose. 

We use \bonnsai in combination with precise measurements of stellar
masses and radii of Milky Way binaries \citep{2010A&ARv..18...67T}
to test the Milky Way stellar models of \citet{2011A&A...530A.115B}.
The stellar masses and radii of the Milky Way binary components are
determined from observed radial-velocity curves and light curves.
We call masses and radii determined in this way dynamical masses and
dynamical radii. The surface gravities, $\log g$, follow directly
from the measured dynamical masses and radii. Additionally, the stellar
bolometric luminosities follow from the Stefan-Boltzmann law if the
effective temperatures are known as well. The latter typically rely
on multi-band photometry and calibrations, individual spectra or a
comparison of the spectral energy distributions obtained by narrow
band filters with stellar atmosphere models (spectro-photometry).

\begin{table*}
\caption{Observed stellar parameters of those Milky Way binaries in the sample
of \citet{2010A&ARv..18...67T} that we use to test \bonnsai and the
stellar models of \citet{2011A&A...530A.115B}. Listed are the orbital
period $P$ (semi major axis $a$ in parentheses), the dynamical mass
$M_{\mathrm{dyn}}$ and radius $R_{\mathrm{dyn}}$, the projected
rotational velocity $v\sin i$, the spectral type SpT, the effective
temperature $T_{\mathrm{eff,obs}}$, the surface gravity $\log g_{\mathrm{obs}}$
and the bolometric luminosity $\log L_{\mathrm{obs}}$.}

\centering
\begin{tabular}{lcccccccccc}
\toprule
Binary & P (days) & & SpT & $M_{\mathrm{dyn}}$ ($\msun$) & $R_{\mathrm{dyn}}$ ($\rsun$) & $v\sin i$ ($\mathrm{km}\,\mathrm{s}^{-1}$) & $T_{\mathrm{eff,obs}}$ (K) & $\log g_{\mathrm{obs}}$ (cgs) & $\log L_{\mathrm{obs}}$ ($\lsun$) \tabularnewline
\midrule 
\midrule 
V3903 Sgr & 1.74 & A & O7V & 27.27$\pm$0.55 & 8.088$\pm$0.086 & 230$\pm$23 & 38,000$\pm$1900 & 4.058$\pm$0.016 & 5.088$\pm$0.087\tabularnewline
 & (21.9 $\rsun$) & B & O9V & 19.01$\pm$0.44 & 6.125$\pm$0.060 & 170$\pm$17 & 34,100$\pm$1700 & 4.143$\pm$0.013 & 4.658$\pm$0.088\tabularnewline
\midrule 
EM Car & 3.41 & A & O8V & 22.83$\pm$0.32 & 9.350$\pm$0.170 & 150$\pm$20 & 34,000$\pm$2000 & 3.855$\pm$0.016 & 5.021$\pm$0.104\tabularnewline
 & (33.7 $\rsun$) & B & O8V & 21.38$\pm$0.33 & 8.350$\pm$0.160 & 130$\pm$15 & 34,000$\pm$2000 & 3.925$\pm$0.016 & 4.922$\pm$0.104\tabularnewline
\midrule 
V1034 Sco & 2.44 & A & O9V & 17.21$\pm$0.46 & 7.507$\pm$0.081 & $\dots$ & 33,200$\pm$900 & 3.923$\pm$0.008 & 4.789$\pm$0.048\tabularnewline
 & (22.8 $\rsun$) & B & B1.5V & 9.59$\pm$0.27 & 4.217$\pm$0.089 & $\dots$ & 26,330$\pm$900 & 4.170$\pm$0.013 & 3.885$\pm$0.062\tabularnewline
\midrule 
V478 Cyg & 2.88 & A & O9.5V & 16.62$\pm$0.33 & 7.426$\pm$0.072 & $\dots$ & 30,479$\pm$1000 & 3.917$\pm$0.007 & 4.631$\pm$0.058\tabularnewline
 & (27.3 $\rsun$) & B & O9.5V & 16.27$\pm$0.33 & 7.426$\pm$0.072 & $\dots$ & 30,549$\pm$1000 & 3.908$\pm$0.008 & 4.635$\pm$0.057\tabularnewline
\midrule 
AH Cep & 1.77 & A & B0.5Vn & 15.26$\pm$0.35 & 6.346$\pm$0.071 & 185$\pm$30 & 29,900$\pm$1000 & 4.017$\pm$0.009 & 4.461$\pm$0.059\tabularnewline
 & (18.9 $\rsun$) & B & B0.5Vn & 13.44$\pm$0.25 & 5.836$\pm$0.085 & 185$\pm$30 & 28,600$\pm$1000 & 4.034$\pm$0.012 & 4.311$\pm$0.062\tabularnewline
\midrule 
V578 Mon & 2.41 & A & B1V & 14.50$\pm$0.12 & 5.149$\pm$0.091 & 117$\pm$5 & 30,000$\pm$740 & 4.176$\pm$0.015 & 4.285$\pm$0.045\tabularnewline
 & (22.0 $\rsun$) & B & B2V & 10.26$\pm$0.08 & 4.210$\pm$0.100 & 94$\pm$4 & 26,400$\pm$600 & 4.200$\pm$0.021 & 3.888$\pm$0.045\tabularnewline
\midrule 
V453 Cyg & 3.89 & A & B0.4IV & 13.82$\pm$0.35 & 8.445$\pm$0.068 & 109$\pm$3 & 27,800$\pm$400 & 3.725$\pm$0.006 & 4.583$\pm$0.026\tabularnewline
 & (30.2 $\rsun$) & B & B0.7IV & 10.64$\pm$0.22 & 5.420$\pm$0.068 & 98$\pm$5 & 26,200$\pm$500 & 3.997$\pm$0.010 & 4.094$\pm$0.035\tabularnewline
\midrule 
CW Cep & 2.73 & A & B0.5V & 13.05$\pm$0.20 & 5.640$\pm$0.120 & $\dots$ & 28,300$\pm$1000 & 4.050$\pm$0.019 & 4.263$\pm$0.064\tabularnewline
 & (24.0 $\rsun$) & B & B0.5V & 11.91$\pm$0.20 & 5.140$\pm$0.120 & $\dots$ & 27,700$\pm$1000 & 4.092$\pm$0.021 & 4.145$\pm$0.067\tabularnewline
\midrule 
DW Car & 1.33 & A & B1V & 11.34$\pm$0.18 & 4.561$\pm$0.050 & 182$\pm$3 & 27,900$\pm$1000 & 4.175$\pm$0.009 & 4.054$\pm$0.063\tabularnewline
 & (14.3 $\rsun$) & B & B1V & 10.63$\pm$0.20 & 4.299$\pm$0.058 & 177$\pm$3 & 26,500$\pm$1000 & 4.198$\pm$0.011 & 3.913$\pm$0.067\tabularnewline
\midrule 
QX Car & 4.48 & A & B2V & 9.25$\pm$0.12 & 4.291$\pm$0.091 & 120$\pm$10 & 23,800$\pm$500 & 4.139$\pm$0.018 & 3.725$\pm$0.041\tabularnewline
 & (29.8 $\rsun$) & B & B2V & 8.46$\pm$0.12 & 4.053$\pm$0.091 & 110$\pm$10 & 22,600$\pm$500 & 4.150$\pm$0.019 & 3.585$\pm$0.043\tabularnewline
\midrule 
V1388 Ori & 2.19 & A & B2.5IV-V & 7.42$\pm$0.16 & 5.600$\pm$0.080 & 125$\pm$10 & 20,500$\pm$500 & 3.812$\pm$0.016 & 3.697$\pm$0.044\tabularnewline
 & (16.5 $\rsun$) & B & B3V & 5.16$\pm$0.06 & 3.760$\pm$0.060 & 75$\pm$15 & 18,500$\pm$500 & 4.000$\pm$0.015 & 3.172$\pm$0.049\tabularnewline
\midrule 
V539 Ara & 3.17 & A & B3V & 6.24$\pm$0.07 & 4.516$\pm$0.084 & 75$\pm$8 & 18,100$\pm$500 & 3.924$\pm$0.016 & 3.293$\pm$0.051\tabularnewline
 & (20.5 $\rsun$) & B & B4V & 5.31$\pm$0.06 & 3.428$\pm$0.083 & 48$\pm$5 & 17,100$\pm$500 & 4.093$\pm$0.021 & 2.955$\pm$0.055\tabularnewline
\midrule 
CV Vel & 6.89 & A & B2.5V & 6.09$\pm$0.04 & 4.089$\pm$0.036 & 19$\pm$1 & 18,100$\pm$500 & 3.999$\pm$0.008 & 3.207$\pm$0.049\tabularnewline
 & (35.0 $\rsun$) & B & B2.5V & 5.98$\pm$0.04 & 3.950$\pm$0.036 & 31$\pm$2 & 17,900$\pm$500 & 4.022$\pm$0.008 & 3.158$\pm$0.049\tabularnewline
\midrule 
AG Per & 2.03 & A & B3.4V & 5.35$\pm$0.16 & 2.995$\pm$0.071 & 94$\pm$23 & 18,200$\pm$800 & 4.213$\pm$0.020 & 2.946$\pm$0.079\tabularnewline
 & (14.7 $\rsun$) & B & B3.5V & 4.89$\pm$0.13 & 2.605$\pm$0.070 & 70$\pm$9 & 17,400$\pm$800 & 4.296$\pm$0.023 & 2.747$\pm$0.083\tabularnewline
\midrule 
U Oph & 1.68 & A & B5V & 5.27$\pm$0.09 & 3.484$\pm$0.021 & 125$\pm$5 & 16,440$\pm$250 & 4.076$\pm$0.004 & 2.901$\pm$0.027\tabularnewline
 & (12.8 $\rsun$) & B & B6V & 4.74$\pm$0.07 & 3.110$\pm$0.034 & 115$\pm$5 & 15,590$\pm$250 & 4.128$\pm$0.009 & 2.710$\pm$0.029\tabularnewline
\midrule 
DI Her & 10.55 & A & B5V & 5.17$\pm$0.11 & 2.681$\pm$0.046 & $\dots$ & 17,000$\pm$800 & 4.295$\pm$0.015 & 2.732$\pm$0.083\tabularnewline
 & (43.2 $\rsun$) & B & B5V & 4.52$\pm$0.07 & 2.478$\pm$0.046 & $\dots$ & 15,100$\pm$700 & 4.305$\pm$0.015 & 2.457$\pm$0.082\tabularnewline
\midrule 
V760 Sco & 1.73 & A & B4V & 4.97$\pm$0.09 & 3.015$\pm$0.066 & 95$\pm$10 & 16,900$\pm$500 & 4.176$\pm$0.019 & 2.823$\pm$0.055\tabularnewline
 & (12.9 $\rsun$) & B & B4V & 4.61$\pm$0.07 & 2.641$\pm$0.066 & 85$\pm$10 & 16,300$\pm$500 & 4.258$\pm$0.021 & 2.645$\pm$0.058\tabularnewline
\midrule 
MU Cas & 9.65 & A & B5V & 4.66$\pm$0.10 & 4.195$\pm$0.058 & 21$\pm$2 & 14,750$\pm$800 & 3.861$\pm$0.012 & 2.874$\pm$0.096\tabularnewline
 & (40.0 $\rsun$) & B & B5V & 4.58$\pm$0.09 & 3.670$\pm$0.057 & 22$\pm$2 & 15,100$\pm$800 & 3.969$\pm$0.013 & 2.798$\pm$0.094\tabularnewline
\bottomrule
\end{tabular}

\label{tab:stellar-parameters-torres}
\end{table*}

The \citet{2010A&ARv..18...67T} sample of Milky Way binary stars
is an updated extension of the sample of \citet{1991A&ARv...3...91A}.
All binary stars are analysed homogeneously and dynamical masses and
radii are known to better than 3\%. We work with a subsample of the
\citet{2010A&ARv..18...67T} sample, namely with all stars that are
within the mass range covered by our Milky Way stellar models. We
extend the published stellar models of \citet{2011A&A...530A.115B}
by unpublished ones down to $4\,\msun$ to increase our binary sample
size by three. We describe our method to test the stellar
models in Sec.~\ref{sec:description-of-test}, explain why stellar
rotation needs to be accounted for in Sec.~\ref{sec:importance-of-rotation},
present the results of our test in Sec.~\ref{sec:comparison-of-ages}
and compare effective temperatures and bolometric luminosities predicted
by the models to the observed values in Sec.~\ref{sec:teff-and-logl}.

\subsection{Description of our test}\label{sec:description-of-test}

The primary and secondary stars in binaries can be assumed to be coeval.
We use this condition to test stellar models by comparing the ages
determined individually for the primary and secondary star of each
binary. The ages inferred for the primary and secondary star may deviate
if the physics and calibrations of the stellar models are not accurate.
This or a similar test is often applied to binaries in order to test
stellar evolution and investigate convective core overshooting \citep[see e.g.][and references therein]{1991A&ARv...3...91A,1997MNRAS.285..696S,1997MNRAS.289..869P,2010A&A...516A..42C,2010A&ARv..18...67T,2014AJ....147...36T}.

In case of observed dynamical masses and radii, the stellar ages are
constrained through the time dependence of stellar radii because the
masses of stars in the binary sample of \citet{2010A&ARv..18...67T}
hardly change with time. However, stellar radii, $R$, depend not
only on age but on a variety of parameters,
\begin{equation}
R=R(M,\,\tau,\, v_{\mathrm{rot}},\, Z,\,\alpha_{\mathrm{ML}},\, l_{\mathrm{ov}},\dots),\nonumber
\end{equation}
such as --- in addition to age $\tau$ --- mass $M$, rotational velocity
$v_{\mathrm{rot}}$, metallicity $Z$ and the treatment of convection,
indicated here by the convective mixing length parameter $\alpha_{\mathrm{ML}}$
and the convective core overshooting length $l_{\mathrm{ov}}$. The
mixing length parameter, $\alpha_{\mathrm{ML}}$, plays only a minor
role in the hot stars considered here. Besides the dynamically measured
masses and radii, we use, if available, the observed projected rotational
velocities $v\sin i$ to determine stellar ages (Table~\ref{tab:stellar-parameters-torres}).
The parameters radius, mass and rotational velocity are fixed within
their uncertainties by the observations and the derived stellar ages
depend on the remaining parameters. That is, our test probes the chemical
composition (metallicity) and the implementation and calibration of
rotation and convection in the stellar models. The metallicities are
also known to a certain degree because the binary sample consists
of Milky Way stars.

Because of the precise observations, we use denser stellar model grids
than those available by default in the \bonnsai web-service. With the
higher resolution our resolution test, the $\chi^{2}$-hypothesis
test and our posterior predictive check are passed by all stars.

Our test loses significance if the stars in a binary have very similar
masses and radii because the ages derived for such similar stars have
to be the same no matter which stellar models and calibrations are
used. In the following we therefore cite the mass ratios $q$ of secondary
to primary star to easily spot binaries with similar stellar components. 

We compare the stars in the binaries to stellar evolutionary models
of single stars. This assumption is good as long as the past evolution
of the stars is not influenced significantly by binary interactions.
Past mass transfer episodes are not expected to have occurred in our
binary sample because all the stars are on the main-sequence and presently
do not fill their Roche lobes.
Tidal interactions, however, have influenced
the binaries and affect stellar radii in two ways: (1) tides spin
stars up or down which consequently changes their radii by rotational
mixing and centrifugal forces; (2) tides dissipate energy inside stars,
thereby giving rise to an additional energy source that might increase
stellar radii. All binary stars in \citet{2010A&ARv..18...67T} whose
radii are larger than about 25\% of the orbital separation are circularised,
implying that the stellar rotation periods are synchronised with the
orbital periods because synchronisation is expected to precede circularisation.
Once the binaries are synchronised and circularised, no torques act
on the stars and tides no longer influence the evolution until stellar
evolution either changes the spin period of the stars or the orbit
by e.g. wind mass loss such that tides are active again. We have to
keep this in mind when comparing the observations to single star models
that do not take tidal evolution into account.

\subsection{The role of rotation in Milky Way binaries}\label{sec:importance-of-rotation}

The stars with known rotation rates in our sample have projected rotational
velocities $v\sin i$ between $20$ and $230\,\mathrm{km}\,\mathrm{s}^{-1}$.
The associated centrifugal forces lead to larger stellar radii. Within
the Roche model, we approximate by how much stellar radii change as
a function of rotational velocity. The Roche potential $\psi(r,\vartheta)$
of a star with mass $M$ rotating with frequency $\Omega=2\pi/P_{\mathrm{s}}$
($P_{\mathrm{s}}$ being the rotation period) is given by
\begin{equation}
\psi(r,\vartheta)=-\frac{Gm_{r}}{r}-\frac{1}{2}\Omega^{2}r^{2}\sin\vartheta,\label{eq:roche-potential}
\end{equation}
where $r$ is the radial distance and $\vartheta$ the polar angle,
$G$ is the gravitational constant and $m_{r}$ is the mass within
radius $r$. The Roche potential evaluated at the stellar pole ($r=R_{p}$,
$\vartheta=0$) and at the stellar equator ($r=R_{e}$, $\vartheta=\pi/2$)
are equal because the stellar surface is on an equipotential. Equating
$\psi(R,\;0)=\psi(R,\;\pi/2)$, introducing the break-up or critical
rotational frequency $\Omega_{\mathrm{crit}}=\sqrt{GM/R_{e}^{3}}$
(i.e. Keplerian frequency at the equator) and $\Gamma=\Omega/\Omega_{\mathrm{crit}}$,
we have for the relative difference of the equatorial and polar radius
\begin{equation}
\frac{\Delta R}{R}\equiv\frac{R_{e}-R_{p}}{R_{p}}=\frac{\Gamma^{2}}{2}.\label{eq:rel-radius-diff-rotation}
\end{equation}
The polar radius $R_{p}$ is not explicitly affected by rotation and
hence it can be viewed as the radius of a star in absence of rotation
(centrifugal forces do not act in the direction of the rotation axis).
Hence, Eq.~\ref{eq:rel-radius-diff-rotation} provides an estimate
of the relative increase in equatorial radius of stars rotating with
a fraction $\Gamma$ of critical rotation. The equatorial radius increases
by 3\% if the star rotates with about 25\% of critical rotation.

\begin{figure}
\begin{centering}
\includegraphics[width=8.5cm]{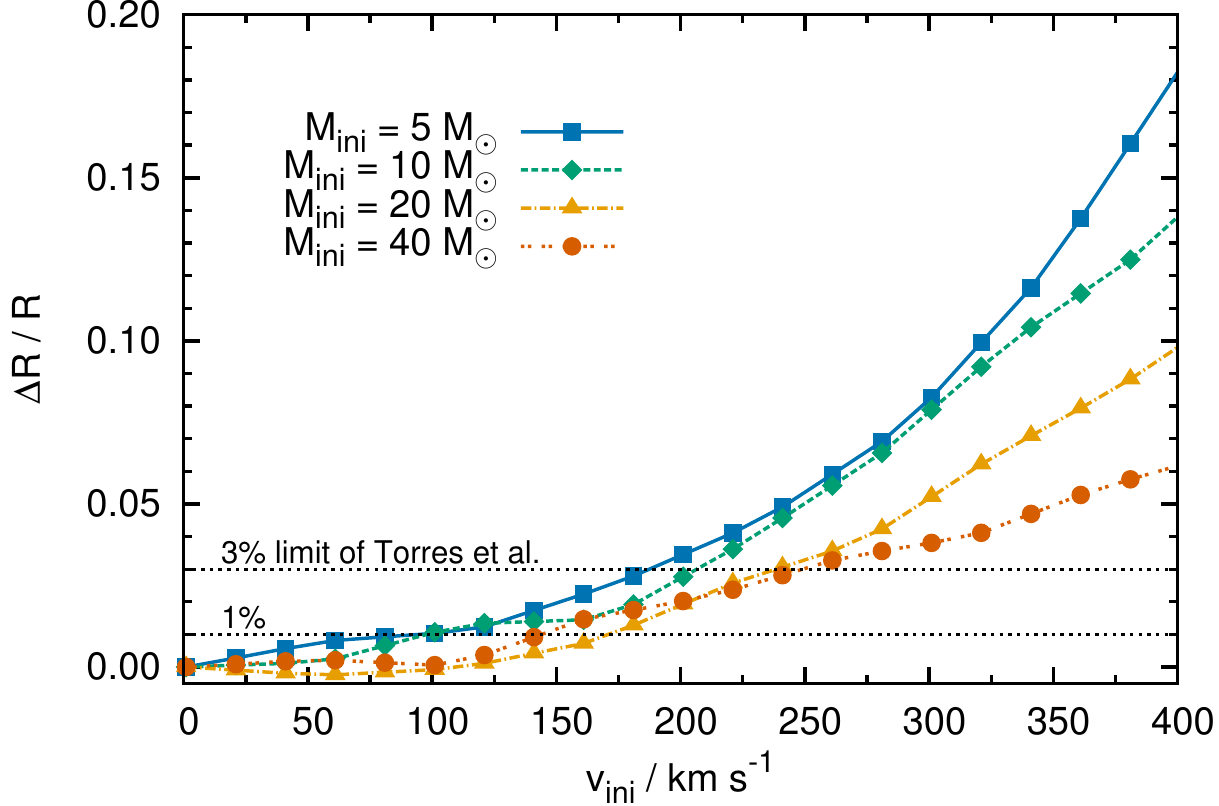}
\par\end{centering}

\caption{Relative radius difference of zero-age, rotating and non-rotating
stellar models from \citet{2011A&A...530A.115B} as a function of
initial rotational velocity $v_{\mathrm{ini}}$ and initial stellar
mass $M_{\mathrm{ini}}$.}

\label{fig:rel-radius-diff-rotation}
\end{figure}

In Fig.~\ref{fig:rel-radius-diff-rotation} we show the increase
of the equatorial stellar radii of rotating zero-age main-sequence
stars compared to non-rotating stars ($\Delta R/R\equiv\left[R_{\mathrm{rot}}-R_{\mathrm{non-rot}}\right]/R_{\mathrm{non-rot}}$)
as a function of initial rotational velocity $v_{\mathrm{ini}}$ and
initial stellar mass $M_{\mathrm{ini}}$. 
Stars that have masses $\leq 10\,\msun$ and rotate with about $100\,\mathrm{km}\,\mathrm{s}^{-1}$ have radii
increased by about 1\%. The radii are increased by more than 3\% if stars rotate faster than
about $200\,\mathrm{km}\,\mathrm{s}^{-1}$.
Given that dynamical radii in \citet{2010A&ARv..18...67T}
are all known to better than 3\% and often to about 1\%, our estimates
show that, at such accuracies, rotation has to be considered to accurately
test the stellar models.

\subsection{The ages of primary and secondary stars}\label{sec:comparison-of-ages}

We determine stellar parameters of the binary stars from dynamical
masses, dynamical radii and projected rotational velocities (Table~\ref{tab:stellar-parameters-torres}).
The latter are not available for the binaries V1034~Sco, V478~Cyg,
CW~Cep and DI~Her, so we determine their stellar parameters from
their dynamical masses and radii alone. Our determined initial mass,
age, initial rotational velocity, effective temperature, surface gravity
and luminosity are summarised in Table~\ref{tab:stellar-parameters-bonnsai}
including their (mostly) $1\sigma$ uncertainties (see caption of
Table~\ref{tab:stellar-parameters-bonnsai}). We also indicate the
approximate fractional main-sequence age, $\tau/\tau_{\mathrm{MS}}$,
of the stars in 5\% steps. The main-sequence lifetimes, $\tau_{\mathrm{MS}}$,
of the stellar models are determined from the obtained initial mass
and initial rotational velocity. 

\begin{table*}
\caption{Evolutionary parameters of the Milky Way binaries from Table~\ref{tab:stellar-parameters-torres}
as determined with \bonnsai. We provide the initial stellar mass $M_{\mathrm{ini}}$,
the stellar age $\tau$, the initial rotational velocity $v_{\mathrm{ini}}$,
the fractional main-sequence age $\tau/\tau_{\mathrm{MS}}$, the effective
temperature $T_{\mathrm{eff,theo}}$, the surface gravity $\log g_{\mathrm{theo}}$
and the bolometric luminosity $\log L_{\mathrm{theo}}$, including
their (mostly) $1\sigma$ (68.3\%) confidence intervals. Note that
the confidence level of the surface gravity of CV~Vel~A is 71.5\%,
that of the initial rotational velocity of V1034~Sco~B is 76.3\%
and that of the effective temperatures of CW~Cep~A and CW~Cep~B
are 81.8\% and 87.0\%, respectively (because of broad posterior distributions
that do not allow us to compute smaller confidence intervals). }

\centering
\begin{tabular}{lcccccccc}
\toprule 
Binary &  & $M_{\mathrm{ini}}$ ($\msun$) & $\tau$ (Myr) & $v_{\mathrm{ini}}$ ($\mathrm{km}\,\mathrm{s}^{-1}$) & $\tau/\tau_\mathrm{MS}$ & $T_{\mathrm{eff,theo}}$ (K) & $\log g_{\mathrm{theo}}$ (cgs) & $\log L_{\mathrm{theo}}$ ($\lsun$) \tabularnewline
\midrule
\midrule
V3903 Sgr & A & $27.70^{+0.60}_{-0.62}$ & $2.0^{+0.3}_{-0.2}$ & $240.0^{+101.8}_{-36.1}$ & 30\% & $37,475^{+402}_{-424}$ & $4.08^{+0.02}_{-0.04}$ & $5.08^{+0.03}_{-0.04}$\tabularnewline
 & B & $19.02^{+0.44}_{-0.44}$ & $1.9^{+0.4}_{-0.4}$ & $180.0^{+116.4}_{-40.4}$ & 20\% & $33,275^{+506}_{-590}$ & $4.12^{+0.04}_{-0.03}$ & $4.62^{+0.03}_{-0.05}$\tabularnewline
\midrule
EM Car & A & $23.28^{+0.34}_{-0.36}$ & $4.3^{+0.2}_{-0.2}$ & $160.0^{+85.5}_{-34.6}$ & 60\% & $33,425^{+327}_{-331}$ & $3.88^{+0.02}_{-0.05}$ & $4.97^{+0.04}_{-0.03}$\tabularnewline
 & B & $21.66^{+0.35}_{-0.35}$ & $4.2^{+0.2}_{-0.2}$ & $140.0^{+85.4}_{-37.5}$ & 55\% & $33,375^{+319}_{-386}$ & $3.92^{+0.03}_{-0.03}$ & $4.88^{+0.04}_{-0.03}$\tabularnewline
\midrule
V1034 Sco & A & $17.30^{+0.49}_{-0.46}$ & $5.7^{+0.6}_{-0.5}$ & $520.0^{+4.9}_{-350.1}$ & 45\%--60\% & $30,375^{+581}_{-611}$ & $3.92^{+0.03}_{-0.03}$ & $4.62^{+0.05}_{-0.04}$\tabularnewline
 & B & $9.56^{+0.26}_{-0.28}$ & $5.7^{+2.0}_{-2.6}$ & $420.0^{+48.9}_{-318.7}$ & 20\% & $24,275^{+536}_{-1003}$ & $4.17^{+0.03}_{-0.04}$ & $3.73^{+0.06}_{-0.06}$\tabularnewline
\midrule
V478 Cyg & A & $16.70^{+0.36}_{-0.32}$ & $6.1^{+0.5}_{-0.5}$ & $520.0^{+6.7}_{-346.3}$ & 45\%--60\% & $29,875^{+488}_{-522}$ & $3.92^{+0.02}_{-0.03}$ & $4.58^{+0.06}_{-0.02}$\tabularnewline
 & B & $16.36^{+0.34}_{-0.34}$ & $6.4^{+0.5}_{-0.5}$ & $520.0^{+7.5}_{-346.7}$ & 50\%--60\% & $29,475^{+518}_{-499}$ & $3.92^{+0.02}_{-0.04}$ & $4.58^{+0.04}_{-0.04}$\tabularnewline
\midrule
AH Cep & A & $15.28^{+0.37}_{-0.35}$ & $5.3^{+0.5}_{-0.4}$ & $200.0^{+90.5}_{-52.8}$ & 45\% & $29,475^{+478}_{-447}$ & $4.03^{+0.03}_{-0.03}$ & $4.42^{+0.05}_{-0.03}$\tabularnewline
 & B & $13.46^{+0.24}_{-0.26}$ & $6.2^{+0.5}_{-0.5}$ & $200.0^{+91.5}_{-52.7}$ & 45\% & $27,975^{+386}_{-512}$ & $4.03^{+0.03}_{-0.03}$ & $4.28^{+0.03}_{-0.05}$\tabularnewline
\midrule
V578 Mon & A & $14.50^{+0.13}_{-0.11}$ & $2.3^{+0.6}_{-0.5}$ & $120.0^{+87.3}_{-33.9}$ & 20\% & $29,975^{+244}_{-270}$ & $4.17^{+0.03}_{-0.03}$ & $4.28^{+0.03}_{-0.03}$\tabularnewline
 & B & $10.26^{+0.09}_{-0.08}$ & $4.0^{+1.1}_{-1.1}$ & $100.0^{+66.0}_{-34.4}$ & 20\% & $25,475^{+218}_{-208}$ & $4.17^{+0.06}_{-0.01}$ & $3.83^{+0.03}_{-0.03}$\tabularnewline
\midrule
V453 Cyg & A & $13.84^{+0.37}_{-0.34}$ & $10.2^{+0.5}_{-0.6}$ & $120.0^{+71.0}_{-37.2}$ & 80\% & $26,025^{+449}_{-492}$ & $3.73^{+0.02}_{-0.02}$ & $4.47^{+0.03}_{-0.05}$\tabularnewline
 & B & $10.64^{+0.22}_{-0.22}$ & $10.9^{+0.8}_{-0.9}$ & $100.0^{+78.5}_{-31.4}$ & 55\% & $24,925^{+368}_{-439}$ & $3.98^{+0.05}_{-0.02}$ & $4.03^{+0.02}_{-0.06}$\tabularnewline
\midrule
CW Cep & A & $13.06^{+0.21}_{-0.20}$ & $6.2^{+0.6}_{-0.7}$ & $520.0^{+30.0}_{-327.0}$ & 35\%--40\% & $27,825^{+308}_{-1,285}$ & $4.08^{+0.01}_{-0.06}$ & $4.22^{+0.03}_{-0.07}$\tabularnewline
 & B & $11.92^{+0.20}_{-0.20}$ & $5.9^{+1.1}_{-1.2}$ & $520.0^{+30.0}_{-305.0}$ & 30\%--35\% & $25,475^{+1,746}_{-373}$ & $4.08^{+0.05}_{-0.02}$ & $4.08^{+0.03}_{-0.09}$\tabularnewline
\midrule
DW Car & A & $11.34^{+0.17}_{-0.19}$ & $3.7^{+0.7}_{-0.8}$ & $200.0^{+83.6}_{-47.0}$ & 20\% & $26,475^{+333}_{-465}$ & $4.17^{+0.02}_{-0.02}$ & $3.98^{+0.02}_{-0.06}$\tabularnewline
 & B & $10.62^{+0.19}_{-0.21}$ & $3.3^{+1.0}_{-1.0}$ & $180.0^{+87.5}_{-30.7}$ & 15\% & $25,675^{+396}_{-412}$ & $4.17^{+0.05}_{-0.02}$ & $3.88^{+0.02}_{-0.06}$\tabularnewline
\midrule
QX Car & A & $9.24^{+0.13}_{-0.11}$ & $8.4^{+1.2}_{-1.5}$ & $140.0^{+71.8}_{-49.6}$ & 30\% & $23,825^{+293}_{-327}$ & $4.12^{+0.04}_{-0.03}$ & $3.73^{+0.03}_{-0.04}$\tabularnewline
 & B & $8.46^{+0.12}_{-0.12}$ & $9.6^{+1.6}_{-1.7}$ & $120.0^{+75.5}_{-37.5}$ & 30\% & $22,775^{+267}_{-345}$ & $4.12^{+0.06}_{-0.01}$ & $3.58^{+0.06}_{-0.03}$\tabularnewline
\midrule
V1388 Ori & A & $7.44^{+0.16}_{-0.17}$ & $29.4^{+1.9}_{-1.7}$ & $140.0^{+95.7}_{-36.0}$ & 75\% & $19,275^{+383}_{-446}$ & $3.83^{+0.03}_{-0.04}$ & $3.58^{+0.06}_{-0.04}$\tabularnewline
 & B & $5.16^{+0.07}_{-0.06}$ & $51.6^{+3.0}_{-3.3}$ & $80.0^{+76.3}_{-33.6}$ & 60\% & $16,475^{+188}_{-249}$ & $3.98^{+0.06}_{-0.01}$ & $2.98^{+0.03}_{-0.04}$\tabularnewline
\midrule
V539 Ara & A & $6.26^{+0.06}_{-0.08}$ & $38.0^{+1.8}_{-1.8}$ & $80.0^{+78.4}_{-33.3}$ & 65\% & $18,125^{+206}_{-256}$ & $3.92^{+0.03}_{-0.03}$ & $3.27^{+0.05}_{-0.03}$\tabularnewline
 & B & $5.32^{+0.06}_{-0.06}$ & $38.4^{+3.5}_{-3.8}$ & $60.0^{+49.2}_{-40.9}$ & 50\% & $17,225^{+177}_{-227}$ & $4.08^{+0.05}_{-0.02}$ & $2.98^{+0.03}_{-0.04}$\tabularnewline
\midrule
CV Vel & A & $6.10^{+0.04}_{-0.05}$ & $36.0^{+1.0}_{-1.2}$ & $20.0^{+39.8}_{-18.2}$ & 60\% & $18,275^{+118}_{-119}$ & $4.03^{+0.03}_{-0.04}$ & $3.23^{+0.02}_{-0.03}$\tabularnewline
 & B & $5.98^{+0.05}_{-0.03}$ & $35.6^{+1.0}_{-1.2}$ & $40.0^{+37.9}_{-33.0}$ & 60\% & $18,175^{+121}_{-94}$ & $4.03^{+0.02}_{-0.02}$ & $3.17^{+0.02}_{-0.02}$\tabularnewline
\midrule
AG Per & A & $5.34^{+0.15}_{-0.17}$ & $19.4^{+5.8}_{-5.7}$ & $100.0^{+69.9}_{-35.5}$ & 25\% & $17,625^{+396}_{-442}$ & $4.22^{+0.03}_{-0.05}$ & $2.88^{+0.06}_{-0.04}$\tabularnewline
 & B & $4.86^{+0.14}_{-0.12}$ & $9.6^{+5.7}_{-5.2}$ & $80.0^{+45.5}_{-35.8}$ & 10\% & $17,025^{+347}_{-378}$ & $4.28^{+0.05}_{-0.02}$ & $2.73^{+0.03}_{-0.07}$\tabularnewline
\midrule
U Oph & A & $5.28^{+0.09}_{-0.10}$ & $39.8^{+3.7}_{-4.1}$ & $140.0^{+82.0}_{-39.0}$ & 50\% & $16,875^{+324}_{-286}$ & $4.08^{+0.02}_{-0.02}$ & $2.92^{+0.07}_{-0.02}$\tabularnewline
 & B & $4.74^{+0.07}_{-0.08}$ & $43.2^{+4.6}_{-5.0}$ & $120.0^{+85.3}_{-31.6}$ & 40\% & $16,025^{+255}_{-281}$ & $4.12^{+0.02}_{-0.02}$ & $2.77^{+0.02}_{-0.06}$\tabularnewline
\midrule
DI Her & A & $5.14^{+0.12}_{-0.10}$ & $6.8^{+3.9}_{-4.2}$ & $60.0^{+155.2}_{-45.0}$ & 10\% & $17,575^{+302}_{-375}$ & $4.28^{+0.04}_{-0.03}$ & $2.77^{+0.05}_{-0.03}$\tabularnewline
 & B & $4.52^{+0.06}_{-0.07}$ & $8.7^{+4.5}_{-5.6}$ & $40.0^{+162.2}_{-25.6}$ & 5\% & $16,225^{+243}_{-255}$ & $4.33^{+0.02}_{-0.05}$ & $2.58^{+0.04}_{-0.03}$\tabularnewline
\midrule
V760 Sco & A & $4.96^{+0.10}_{-0.09}$ & $31.1^{+5.1}_{-5.6}$ & $100.0^{+72.1}_{-32.1}$ & 35\% & $16,725^{+288}_{-287}$ & $4.17^{+0.03}_{-0.03}$ & $2.83^{+0.02}_{-0.06}$\tabularnewline
 & B & $4.60^{+0.08}_{-0.07}$ & $19.5^{+6.0}_{-6.4}$ & $100.0^{+50.9}_{-40.5}$ & 15\% & $16,225^{+275}_{-204}$ & $4.28^{+0.02}_{-0.05}$ & $2.62^{+0.05}_{-0.03}$\tabularnewline
\midrule
MU Cas & A & $4.66^{+0.09}_{-0.10}$ & $81.2^{+5.3}_{-4.8}$ & $20.0^{+45.7}_{-16.9}$ & 70\% & $14,725^{+269}_{-276}$ & $3.88^{+0.03}_{-0.04}$ & $2.88^{+0.04}_{-0.05}$\tabularnewline
 & B & $4.58^{+0.08}_{-0.10}$ & $73.3^{+5.2}_{-4.7}$ & $20.0^{+45.5}_{-16.6}$ & 60\% & $15,075^{+271}_{-244}$ & $3.98^{+0.03}_{-0.03}$ & $2.77^{+0.06}_{-0.02}$\tabularnewline
\bottomrule 
\end{tabular}

\label{tab:stellar-parameters-bonnsai}
\end{table*}

\begin{figure}
\begin{centering}
\includegraphics[width=9cm]{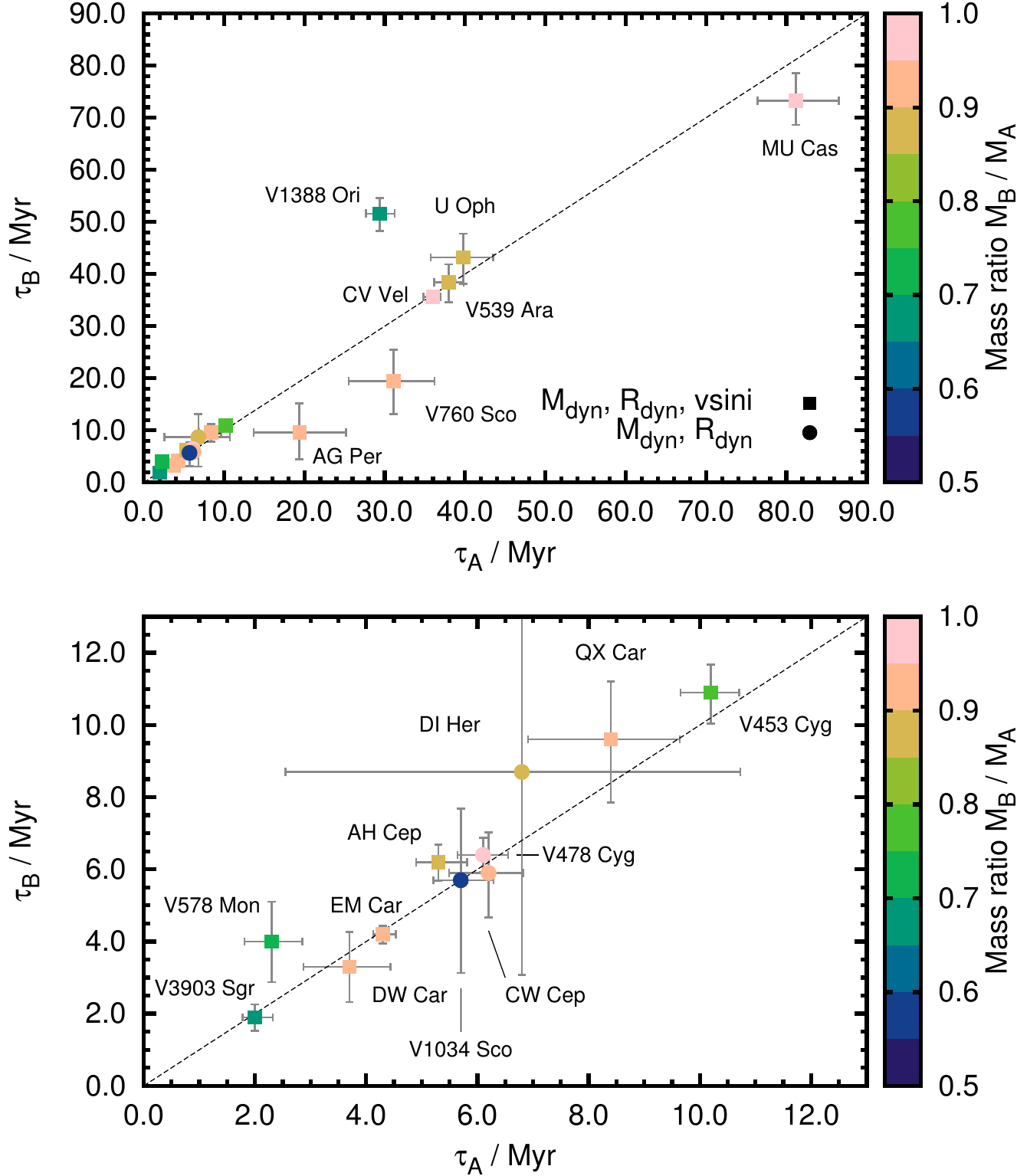}
\par\end{centering}

\caption{Ages of primary ($\tau_{\mathrm{A}}$) and secondary ($\tau_{\mathrm{B}}$)
stars determined from dynamical masses, dynamical radii and (if available)
projected rotational velocities $v\sin i$. The colour coding indicates
the mass ratios of secondary ($M_{\mathrm{B}}$) to primary ($M_{\mathrm{A}}$)
stars. Squares and filled circles show binaries for which $v\sin i$
measurements are available and are lacking to derive stellar
ages, respectively. The bottom panel is a zoom into the age range $0\text{--}12.5\,\mathrm{Myr}$
of the top panel.}

\label{fig:ages}
\end{figure}

The ages $\tau_{\mathrm{A}}$ and $\tau_{\mathrm{B}}$ of the primary
and secondary star of binaries from Table~\ref{tab:stellar-parameters-bonnsai}
are plotted against each other in Fig.~\ref{fig:ages}. If the stellar
models were a perfect representation of the observed stars, the ages
of the binary components should agree within their $1\sigma$ uncertainties
in $68.3\%$, i.e. in 12--13 of the 18 cases. A first inspection reveals
that the ages of the binary components agree well within their uncertainties,
except for one $>6\sigma$ outlier, V1388~Ori. 
We do not have an explanation for the discrepant ages of V1388~Ori. Perhaps
the metallicity of this star differs from solar, which may induce
an age difference. Alternatively, the formation history of the system
may be peculiar.

Stellar radii grow only slowly in the beginning of a stellar life
but more quickly when approaching the end of the main-sequence evolution.
For example, a non-rotating $10\,\msun$ Milky Way star of \citet{2011A&A...530A.115B}
has increased its radius by about 40\% at half of its main-sequence
life but by about 230\% towards its end. Hence, the accuracy with
which ages can be determined from dynamical masses and radii depends
strongly on the fractional main-sequence age of stars and becomes
better the more evolved a star is. Because of this the age of CV~Vel
($\tau/\tau_{\mathrm{MS}}\approx60\%$) is determined to about 3\% whereas
the age of DI~Her ($\tau/\tau_{\mathrm{MS}}\approx5\text{--}10\%$)
is only determined to about 50--60\%, despite an accuracy of 1--2\%
in stellar masses and radii in both cases. 

We quantify the age differences, $\Delta\tau=\tau_{\mathrm{A}}-\tau_{\mathrm{B}}$,
of the primary and secondary stars and evaluate whether the age differences
deviate significantly from zero. As noted in Sec.~\ref{sec:description-of-test},
our test loses significance if both stars in a binary are too similar,
therefore we exclude those binaries from our analysis that have mass
ratios larger than 0.97.
This holds for the binaries V478~Cyg, CV~Vel and MU~Cas (the largest
mass ratio of the remaining binaries, namely that of EM~Car and DW~Car,
is 0.94). We also exclude the outlier V1388~Ori. The mean, relative
age difference of the remaining 14 binaries is $\left\langle \Delta\tau/\delta\Delta\tau\right\rangle =-0.09\pm0.43$
(95\% CI), where $\delta\Delta\tau$ are the $1\sigma$ uncertainties
of the age differences $\Delta\tau$ and the uncertainty $\pm0.43$
is the 95\% confidence interval of the standard error $\sigma/\sqrt{n}$
with $\sigma=0.82$ being the standard deviation of the relative age
differences and $n=14$ the sample size. Further $\chi^{2}$- and
t-tests confirm, with a confidence level of 95\% (p-values of 0.85
and 0.47, respectively), that the age differences are consistent with
being zero (mean age difference $\left\langle \Delta\tau\right\rangle =0.9\pm2.3\,\mathrm{Myr}$,
95\% CI).

Convective core overshooting influences stellar radii most strongly
towards the end of the main-sequence evolution and has little influence
on unevolved stars. A test with a $5\,\msun$ Milky Way star with
an overshooting of 0.5 pressure scale heights compared to no overshooting
shows that the stellar radii differ by less than 3\%--4\% for fractional
main-sequence ages younger than 30\%--40\%. The maximum radii reached
during the main-sequence evolution at fractional main-sequence ages
of 98\%--99\% differ by 53\%, i.e. the model with overshooting has
a radius larger by a factor of $1/(1-0.53)\approx2.1$. In the following we
assume that stars with fractional main-sequence ages less than 35\%
can be viewed as being unaffected by convective core overshooting.
Hence, by restricting the test to those binaries in which
the primary stars have fractional main-sequence ages younger than
about 35\%, i.e. to the binaries V3903~Sgr, V578~Mon, DW~Car, QX~Car,
AG~Per, DI~Her and V760~Sco, we mainly probe the combination of
metallicity and rotation in our models.
The mean, relative age difference of this subsample of binary
stars is $\left\langle \Delta\tau/\delta\Delta\tau\right\rangle =0.14\pm0.73$
(95\% CI) and $\chi^{2}$- and t-tests confirm, with a 95\% confidence
level (p-values of 0.54 and 0.30, respectively), that the age differences
are consistent with zero (mean age difference $\left\langle \Delta\tau\right\rangle =2.4\pm4.2\,\mathrm{Myr}$,
95\% CI). We find no significant evidence for a correlation between
the inferred age- and observed mass-differences of the binary stars.

Our test with the subsample of binaries that are not expected to
be influenced by convective core overshooting shows that the Milky
Way stellar models of \citet{2011A&A...530A.115B}, with their metallicity
and calibration of rotation, reproduce the massive ($4.5\text{--}28\,\msun$)
Milky Way binaries in the sample of \citet{2010A&ARv..18...67T}.
This might come as a surprise because the models are computed for
a metallicity of $Z=0.0088$, which is small compared to current estimates
of the metallicity of the Sun ($Z_{\odot}=0.014\text{--}0.020$; \citealp{1998SSRv...85..161G,2009ARA&A..47..481A}).
Though the total metallicity in the models of \citet{2011A&A...530A.115B}
is rather low compared to that of the Sun, the iron abundance is not.
The opacities in the models are interpolated linearly in the iron
abundances from standard OPAL opacity tables \citep{1996ApJ...464..943I}.
The iron abundance in the \citet{2011A&A...530A.115B} models is $\log(\mathrm{Fe}/\mathrm{H})+12=7.40$
(Table~\ref{tab:Z-in-brott2011}) which is close to the iron abundance
of $\log(\mathrm{Fe}/\mathrm{H})+12\approx7.50$ in the Sun \citep[e.g.][]{1998SSRv...85..161G,2009ARA&A..47..481A},
hence the structures of stars from \citet{2011A&A...530A.115B} follow
those of stars with a solar-like composition.%

Rotation increases the stellar radii at a level which is comparable
to that of the uncertainties of the radii (Fig.~\ref{fig:rel-radius-diff-rotation})
and thus must be accounted for when testing stellar models (Sec.~\ref{sec:importance-of-rotation}).
Rotating stars are larger than non-rotating stars and hence reach
the observed radii earlier in their evolution. Stellar ages inferred
from rotating stellar models are therefore systematically younger
than those inferred from non-rotating stellar models. To quantify
the systematic age shift, we determine the ages of all 36 stars in
our sample using only the non-rotating Milky Way stellar models of
\citet{2011A&A...530A.115B}. We find a mean, relative age difference
of $\left\langle \left(\tau_{\mathrm{rot}}-\tau_{\mathrm{non-rot}}\right)/\tau_{\mathrm{non-rot}}\right\rangle =-7.4\pm1.2\%$
(95\% CI) where the error is again the 95\% confidence interval of
the standard error. The ages inferred from rotating stellar models
are younger than those from non-rotating models for all stars.

\begin{table}

\caption{Iron abundances, total metallicities $Z_{\mathrm{Brott}}$ as given
in \citet{2011A&A...530A.115B} and the corresponding metallicities
$Z_{\kappa}$ of the opacity tables used in the \citet{2011A&A...530A.115B}
models. }

\begin{centering}
\begin{tabular}{cccc}
\toprule 
 & MW & LMC & SMC\tabularnewline
\midrule
\midrule 
$\log(\mathrm{Fe}/\mathrm{H})+12$ & 7.40 & 7.05 & 6.78\tabularnewline
$Z_{\mathrm{Brott}}$ & 0.0088 & 0.0047 & 0.0021\tabularnewline
$Z_{\kappa}$ & 0.0143 & 0.0065 & 0.0035\tabularnewline
\bottomrule
\end{tabular}
\par\end{centering}

\label{tab:Z-in-brott2011}
\end{table}

\subsection{Effective temperatures and bolometric luminosities}\label{sec:teff-and-logl}

\begin{figure}
\begin{centering}
\includegraphics[width=9cm]{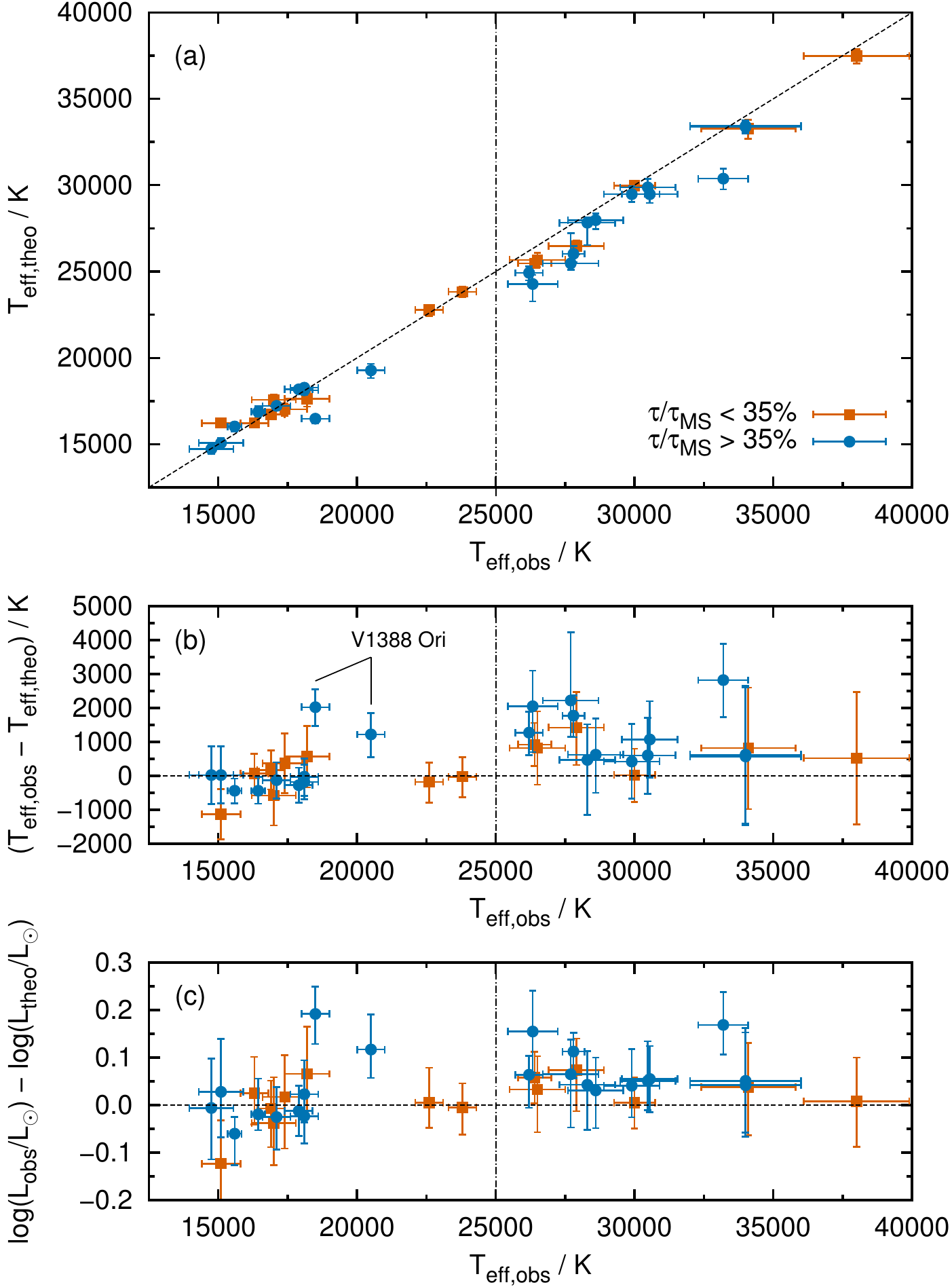}
\par\end{centering}

\caption{Comparison of observed $T_{\mathrm{eff,obs}}$ with predicted effective
temperatures $T_{\mathrm{eff,theo}}$ for stars in our Milky Way binary
sample and its consequence for stellar luminosities (cf. Tables~\ref{tab:stellar-parameters-torres}
and~\ref{tab:stellar-parameters-bonnsai}). Panel (a) shows a direct
comparison of the observed and predicted effective temperatures, panel
(b) the differences of these temperatures $\Delta T_{\mathrm{eff}}=T_{\mathrm{eff,obs}}-T_{\mathrm{eff,theo}}$
as a function of observed effective temperature and panel (c) the
resulting differences of the observed and predicted bolometric luminosities
$\Delta\log L=\log L_{\mathrm{obs}}-\log L_{\mathrm{theo}}$. The
squares show stars with fractional main-sequence ages $\tau/\tau_{\mathrm{MS}}$
younger than 35\% (models not influenced by core overshooting) while
the filled circles represent $\tau/\tau_{\mathrm{MS}}>35\%$ (models
influenced by core overshooting).}

\label{fig:comparison-teff-logl}
\end{figure}

The effective temperatures of stars in the Milky Way binary sample
of \citet{2010A&ARv..18...67T} are independent observables. The effective
temperatures mainly follow from multi-band photometry, calibrations
with respect to spectral types and colours, individual stellar spectra
and spectral energy distributions. Once the effective temperatures
are known, the bolometric luminosities are derived using the Stefan
Boltzmann law and the dynamical radii. Besides determining, e.g.,
stellar ages, we further compute the posterior probability distributions
of the effective temperatures of stars from the measured dynamical
masses, radii and projected rotational velocities. We then compare
the predictions of the stellar models in terms of effective temperatures
and bolometric luminosities to the observations (Fig.~\ref{fig:comparison-teff-logl}).
In the following we exclude V1388~Ori because our stellar models
cannot reproduce this binary.

We find that the observed effective temperatures and hence bolometric
luminosities of our stars are in agreement with the predictions of
the stellar models of \citet{2011A&A...530A.115B} for observed effective
temperatures $T_{\mathrm{eff,obs}}<25,000\,\mathrm{K}$ ($\left\langle \Delta T_{\mathrm{eff}}\right\rangle =133\pm188\,\mathrm{K}$,
95\% CI; $\left\langle \Delta\log L/\lsun\right\rangle =0.01\pm0.02\,\mathrm{dex}$,
95\% CI). However, the observed effective temperatures are on average
hotter by $1062\pm330\,\mathrm{K}$ (95\% CI) and the bolometric luminosities
are consequently larger by $0.06\pm0.02\,\mathrm{dex}$ (95\% CI)
for $T_{\mathrm{eff,obs}}>25,000\,\mathrm{K}$ than predicted by the
stellar models.

The convective core overshooting in our models is not responsible
for the discrepant effective temperatures for $T_{\mathrm{eff,obs}}>25,000\,\mathrm{K}$.
The average difference between observed and predicted effective temperatures
of those stars whose radii are not influenced yet by convective core
overshooting (stars with fractional main-sequence ages $\tau/\tau_{\mathrm{MS}}\leq35\%$)
is $-88\pm377\,\mathrm{K}$ (95\% CI) for $T_{\mathrm{eff,obs}}<25,000\,\mathrm{K}$
and $758\pm371\,\mathrm{K}$ (95\% CI) for $T_{\mathrm{eff,obs}}>25,000\,\mathrm{K}$.
The average differences in effective temperatures of stars expected
to be influenced by core overshooting ($\tau/\tau_{\mathrm{MS}}>35\%$)
are $183\pm496\,\mathrm{K}$ (95\% CI) for $T_{\mathrm{eff,obs}}<25,000\,\mathrm{K}$
and $1213\pm461\,\mathrm{K}$ (95\% CI) for $T_{\mathrm{eff,obs}}>25,000\,\mathrm{K}$.

The cause of the discrepant effective temperatures in stars hotter
than $25,000\,\mathrm{K}$, i.e. earlier than B1--2V or more massive
than about $10\,\msun$, is unknown. We speculate that the discrepancy
is partly explained by calibrations becoming less accurate for more
massive stars. To understand this, we recall how such calibrations
are obtained in practice. In cool stars, fundamental effective temperatures,
i.e. that do not rely on any modelling, can be determined from bolometric
fluxes $F_{\mathrm{bol}}$ that are derived from spectral energy
distributions and from interferometric measurements of stellar angular
diameters $\theta$ ($F_{\mathrm{bol}}=[\theta/2]^{2}\sigma T_{\mathrm{eff}}^{4}$,
where $\sigma$ is the Stefan Boltzmann constant). Such fundamental
measurements of effective temperatures are more complicated and practically
impossible in hot stars because they radiate a substantial fraction
of their light in the ultraviolet (UV) that is difficult to access
observationally (most UV space telescopes cannot observe shortward
of $90\,\mathrm{nm}$ which is the wavelength at which a black body
of $T_{\mathrm{eff}}\approx32,000\,\mathrm{K}$ radiates most of its
energy). Bolometric fluxes of hot stars are therefore difficult to
measure directly and stellar atmosphere models are often used to predict
the flux in the far UV. Instead of this procedure, model atmosphere
computations are applied to infer effective temperatures directly
from spectra and to establish the calibrations with spectral types
and colours. The derived effective temperatures are therefore dependent
on the treatment of, for instance, non-LTE effects, line blanketing,
and stellar winds. Compared to these model atmospheres, the boundary
conditions of the stellar structure models are relatively simple and
this may be a cause of the discrepant effective temperatures.

\section{Conclusions}\label{sec:conclusions}

With the advent of large stellar surveys on powerful telescopes and
advances in analysis techniques of stellar atmospheres more is known
about stars than just their position in the Hertzsprung-Russell (HR)
diagram --- rotation rates, surface gravities and abundances of many
stars are derived. Therefore, the comparison of observations to theoretical
stellar models to infer essential quantities like initial mass or
stellar age and to test stellar models has to be done in a multidimensional
space spanned by all available observables. To that end we develop
\bonnsai\footnote{The \bonnsai web-service is available at 
\href{http://www.astro.uni-bonn.de/stars/bonnsai}{http://www.astro.uni-bonn.de/stars/bonnsai}.}, 
a Bayesian method that allows us to match all available observables
simultaneously to stellar models taking the observed uncertainties
and prior knowledge like mass functions properly into account. Our
method is based on Bayes' theorem from which we determine full (posterior)
probability distributions of the stellar parameters such as initial
mass and stellar age. The probability distributions are analysed to
infer the model parameters including robust uncertainties. \bonnsai
securely identifies cases where the observed stars are not reproduced
by the underlying stellar models through $\chi^{2}$-hypothesis tests
and posterior predictive checks. We test \bonnsai with mock data to
demonstrate its functionality and to show its capabilities.

We apply \bonnsai to the massive star subsample ($\geq4\,\msun$) of the
Milky Way binaries of \citet{2010A&ARv..18...67T}. The masses and
radii of the binaries are known to better than 3\%. For each of the
36 binary components in this sample, we determine the initial masses,
ages, fractional main-sequence ages and initial rotational velocities
from the observed masses, radii and, if available, projected rotational
velocities applying the stellar models of \citet{2011A&A...530A.115B}.%
We find that the Milky Way stellar models of \citet{2011A&A...530A.115B}
result in stellar ages for 17 binaries that are equal within the uncertainties.
There is no statistically significant age difference (95\% confidence
level). We find that \bonnsai, in combination with the Milky Way stellar models
of \citet{2011A&A...530A.115B}, can not fit one binary, V1388~Ori,
for which the ages of both stars differ by $>6\sigma$.

We further compare the effective temperatures predicted by the stellar
models to the observed effective temperatures. The predicted effective
temperatures agree with the observed within their uncertainties for
observed effective temperatures $T_{\mathrm{eff,obs}}\leq25,000\,\mathrm{K}$.
The observed effective temperatures are hotter by $1063\pm330\,\mathrm{K}$
(95\% CI) than the effective temperatures of the models when $T_{\mathrm{eff,obs}}>25,000\,\mathrm{K}$.
The cause of this discrepancy is unknown but may be connected to the
complexities of the atmospheres of hot OB stars. The systematically
hotter temperatures result in stars being brighter by $0.06\pm0.02\,\mathrm{dex}$
(95\% CI) compared to the stellar models.

The \bonnsai approach is flexible and can be easily extended and applied
to different fields in stellar astrophysics. In its current form,
\bonnsai allows us to compare observed stellar surface properties to
models of massive, rotating, main-sequence single stars. Asteroseismology
nowadays offers a look into the interiors of stars. By extending the
stellar models to asteroseismic observables, \bonnsai could make use
of these observables as well to constrain stellar models. Similarly,
\bonnsai can be extended to also match pre main-sequence stars, post
main-sequence stars, low mass stars, binary stars, stars of varying
metallicities etc. to corresponding stellar models. Our statistical
approach also enables the calibration of stellar parameters such as
convective core overshooting including robust uncertainties and to
analyse whole stellar populations to, e.g., unravel initial mass functions
and star formation histories in a statistically sound way while properly
taking observable uncertainties into account.

\begin{acknowledgements}
We thank Selma de Mink, Luca Fossati, Jonathan Mackey, Oscar Ram{\'{\i}}rez-Agudelo,
Hugues Sana, Richard Stancliffe, Frank Tramper and the VFTS consortium
for stimulating discussions, constructive feedback and testing the
\bonnsai web-service. In particular, we thank Claudia McCain for her
help regarding the web-service. We further thank an anonymous referee
for constructive suggestions that helped to improve the presentation
of the paper. FRNS acknowledges the fellowship granted
by the Bonn-Cologne Graduate School of Physics and Astronomy (BCGS)
through the excellence initiative of the German Research Foundation
(DFG). RGI and HHBL thank the Alexander von Humboldt foundation.
\end{acknowledgements}

\bibliographystyle{aa}

\begin{thebibliography}{55}
\expandafter\ifx\csname natexlab\endcsname\relax\def\natexlab#1{#1}\fi

\bibitem[{{Abt} {et~al.}(2002){Abt}, {Levato}, \&
  {Grosso}}]{2002ApJ...573..359A}
{Abt}, H.~A., {Levato}, H., \& {Grosso}, M. 2002, \apj, 573, 359

\bibitem[{{Andersen}(1991)}]{1991A&ARv...3...91A}
{Andersen}, J. 1991, \aapr, 3, 91

\bibitem[{{Asplund} {et~al.}(2009){Asplund}, {Grevesse}, {Sauval}, \&
  {Scott}}]{2009ARA&A..47..481A}
{Asplund}, M., {Grevesse}, N., {Sauval}, A.~J., \& {Scott}, P. 2009, \araa, 47,
  481

\bibitem[{{Barb{\'a}} {et~al.}(2010){Barb{\'a}}, {Gamen}, {Arias}, {Morrell},
  {Ma{\'{\i}}z Apell{\'a}niz}, {Alfaro}, {Walborn}, \&
  {Sota}}]{2010RMxAC..38...30B}
{Barb{\'a}}, R.~H., {Gamen}, R., {Arias}, J.~I., {et~al.} 2010, in Revista
  Mexicana de Astronomia y Astrofisica Conference Series, Vol.~38, Revista
  Mexicana de Astronomia y Astrofisica Conference Series, 30--32

\bibitem[{{Bastian} {et~al.}(2010){Bastian}, {Covey}, \&
  {Meyer}}]{2010ARA&A..48..339B}
{Bastian}, N., {Covey}, K.~R., \& {Meyer}, M.~R. 2010, \araa, 48, 339

\bibitem[{{Bazot} {et~al.}(2012){Bazot}, {Bourguignon}, \&
  {Christensen-Dalsgaard}}]{2012MNRAS.427.1847B}
{Bazot}, M., {Bourguignon}, S., \& {Christensen-Dalsgaard}, J. 2012, \mnras,
  427, 1847

\bibitem[{{Bergemann} {et~al.}(2014){Bergemann}, {Ruchti}, {Serenelli},
  {Feltzing}, {Alves-Brito}, {Asplund}, {Bensby}, {Gruiters}, {Heiter},
  {Hourihane}, {Korn}, {Lind}, {Marino}, {Jofre}, {Nordlander}, {Ryde},
  {Worley}, {Gilmore}, {Randich}, {Ferguson}, {Jeffries}, {Micela},
  {Negueruela}, {Prusti}, {Rix}, {Vallenari}, {Alfaro}, {Allende Prieto},
  {Bragaglia}, {Koposov}, {Lanzafame}, {Pancino}, {Recio-Blanco}, {Smiljanic},
  {Walton}, {Costado}, {Franciosini}, {Hill}, {Lardo}, {de Laverny}, {Magrini},
  {Maiorca}, {Masseron}, {Morbidelli}, {Sacco}, {Kordopatis}, \& {Tautvai{\v
  s}ien{\.e}}}]{2014A&A...565A..89B}
{Bergemann}, M., {Ruchti}, G.~R., {Serenelli}, A., {et~al.} 2014, \aap, 565,
  A89

\bibitem[{{Brott} {et~al.}(2011{\natexlab{a}}){Brott}, {de Mink}, {Cantiello},
  {Langer}, {de Koter}, {Evans}, {Hunter}, {Trundle}, \&
  {Vink}}]{2011A&A...530A.115B}
{Brott}, I., {de Mink}, S.~E., {Cantiello}, M., {et~al.} 2011{\natexlab{a}},
  \aap, 530, A115

\bibitem[{{Brott} {et~al.}(2011{\natexlab{b}}){Brott}, {Evans}, {Hunter}, {de
  Koter}, {Langer}, {Dufton}, {Cantiello}, {Trundle}, {Lennon}, {de Mink},
  {Yoon}, \& {Anders}}]{2011A&A...530A.116B}
{Brott}, I., {Evans}, C.~J., {Hunter}, I., {et~al.} 2011{\natexlab{b}}, \aap,
  530, A116

\bibitem[{{Burnett} \& {Binney}(2010)}]{2010MNRAS.407..339B}
{Burnett}, B. \& {Binney}, J. 2010, \mnras, 407, 339

\bibitem[{{Casagrande} {et~al.}(2011){Casagrande}, {Sch{\"o}nrich}, {Asplund},
  {Cassisi}, {Ram{\'{\i}}rez}, {Mel{\'e}ndez}, {Bensby}, \&
  {Feltzing}}]{2011A&A...530A.138C}
{Casagrande}, L., {Sch{\"o}nrich}, R., {Asplund}, M., {et~al.} 2011, \aap, 530,
  A138

\bibitem[{{Clausen} {et~al.}(2010){Clausen}, {Frandsen}, {Bruntt}, {Olsen},
  {Helt}, {Gregersen}, {Juncher}, \& {Krogstrup}}]{2010A&A...516A..42C}
{Clausen}, J.~V., {Frandsen}, S., {Bruntt}, H., {et~al.} 2010, \aap, 516, A42

\bibitem[{{Conti} \& {Ebbets}(1977)}]{1977ApJ...213..438C}
{Conti}, P.~S. \& {Ebbets}, D. 1977, \apj, 213, 438

\bibitem[{{da Silva} {et~al.}(2006){da Silva}, {Girardi}, {Pasquini},
  {Setiawan}, {von der L{\"u}he}, {de Medeiros}, {Hatzes}, {D{\"o}llinger}, \&
  {Weiss}}]{2006A&A...458..609D}
{da Silva}, L., {Girardi}, L., {Pasquini}, L., {et~al.} 2006, \aap, 458, 609

\bibitem[{{De Gennaro} {et~al.}(2009){De Gennaro}, {von Hippel}, {Jefferys},
  {Stein}, {van Dyk}, \& {Jeffery}}]{2009ApJ...696...12D}
{De Gennaro}, S., {von Hippel}, T., {Jefferys}, W.~H., {et~al.} 2009, \apj,
  696, 12

\bibitem[{{de Mink} {et~al.}(2013){de Mink}, {Langer}, {Izzard}, {Sana}, \& {de
  Koter}}]{2013ApJ...764..166D}
{de Mink}, S.~E., {Langer}, N., {Izzard}, R.~G., {Sana}, H., \& {de Koter}, A.
  2013, \apj, 764, 166

\bibitem[{{Dib}(2014)}]{2014arXiv1405.3287D}
{Dib}, S. 2014, ArXiv e-prints, 1405.3287

\bibitem[{{Dufton} {et~al.}(2013){Dufton}, {Langer}, {Dunstall}, {Evans},
  {Brott}, {de Mink}, {Howarth}, {Kennedy}, {McEvoy}, {Potter},
  {Ram{\'{\i}}rez-Agudelo}, {Sana}, {Sim{\'o}n-D{\'{\i}}az}, {Taylor}, \&
  {Vink}}]{2013A&A...550A.109D}
{Dufton}, P.~L., {Langer}, N., {Dunstall}, P.~R., {et~al.} 2013, \aap, 550,
  A109

\bibitem[{{Evans} {et~al.}(2006){Evans}, {Lennon}, {Smartt}, \&
  {Trundle}}]{2006A&A...456..623E}
{Evans}, C.~J., {Lennon}, D.~J., {Smartt}, S.~J., \& {Trundle}, C. 2006, \aap,
  456, 623

\bibitem[{{Evans} {et~al.}(2005){Evans}, {Smartt}, {Lee}, {Lennon}, {Kaufer},
  {Dufton}, {Trundle}, {Herrero}, {Sim{\'o}n-D{\'{\i}}az}, {de Koter},
  {Hamann}, {Hendry}, {Hunter}, {Irwin}, {Korn}, {Kudritzki}, {Langer},
  {Mokiem}, {Najarro}, {Pauldrach}, {Przybilla}, {Puls}, {Ryans}, {Urbaneja},
  {Venn}, \& {Villamariz}}]{2005A&A...437..467E}
{Evans}, C.~J., {Smartt}, S.~J., {Lee}, J.-K., {et~al.} 2005, \aap, 437, 467

\bibitem[{{Evans} {et~al.}(2011){Evans}, {Taylor}, {H{\'e}nault-Brunet},
  {Sana}, {de Koter}, {Sim{\'o}n-D{\'{\i}}az}, {Carraro}, {Bagnoli}, {Bastian},
  {Bestenlehner}, {Bonanos}, {Bressert}, {Brott}, {Campbell}, {Cantiello},
  {Clark}, {Costa}, {Crowther}, {de Mink}, {Doran}, {Dufton}, {Dunstall},
  {Friedrich}, {Garcia}, {Gieles}, {Gr{\"a}fener}, {Herrero}, {Howarth},
  {Izzard}, {Langer}, {Lennon}, {Ma{\'{\i}}z Apell{\'a}niz}, {Markova},
  {Najarro}, {Puls}, {Ramirez}, {Sab{\'{\i}}n-Sanjuli{\'a}n}, {Smartt},
  {Stroud}, {van Loon}, {Vink}, \& {Walborn}}]{2011A&A...530A.108E}
{Evans}, C.~J., {Taylor}, W.~D., {H{\'e}nault-Brunet}, V., {et~al.} 2011, \aap,
  530, A108

\bibitem[{{Gilmore} {et~al.}(2012){Gilmore}, {Randich}, {Asplund}, {Binney},
  {Bonifacio}, {Drew}, {Feltzing}, {Ferguson}, {Jeffries}, {Micela},
  {Negueruela}, {Prusti}, {Rix}, {Vallenari}, {Alfaro}, {Allende-Prieto},
  {Babusiaux}, {Bensby}, {Blomme}, {Bragaglia}, {Flaccomio}, {Fran{\c c}ois},
  {Irwin}, {Koposov}, {Korn}, {Lanzafame}, {Pancino}, {Paunzen},
  {Recio-Blanco}, {Sacco}, {Smiljanic}, {Van Eck}, \&
  {Walton}}]{2012Msngr.147...25G}
{Gilmore}, G., {Randich}, S., {Asplund}, M., {et~al.} 2012, The Messenger, 147,
  25

\bibitem[{{Grevesse} \& {Sauval}(1998)}]{1998SSRv...85..161G}
{Grevesse}, N. \& {Sauval}, A.~J. 1998, \ssr, 85, 161

\bibitem[{{Gruberbauer} {et~al.}(2012){Gruberbauer}, {Guenther}, \&
  {Kallinger}}]{2012ApJ...749..109G}
{Gruberbauer}, M., {Guenther}, D.~B., \& {Kallinger}, T. 2012, \apj, 749, 109

\bibitem[{{Howarth} {et~al.}(1997){Howarth}, {Siebert}, {Hussain}, \&
  {Prinja}}]{1997MNRAS.284..265H}
{Howarth}, I.~D., {Siebert}, K.~W., {Hussain}, G.~A.~J., \& {Prinja}, R.~K.
  1997, \mnras, 284, 265

\bibitem[{{Huang} {et~al.}(2010){Huang}, {Gies}, \&
  {McSwain}}]{2010ApJ...722..605H}
{Huang}, W., {Gies}, D.~R., \& {McSwain}, M.~V. 2010, \apj, 722, 605

\bibitem[{{Hunter} {et~al.}(2008){Hunter}, {Lennon}, {Dufton}, {Trundle},
  {Sim{\'o}n-D{\'{\i}}az}, {Smartt}, {Ryans}, \& {Evans}}]{2008A&A...479..541H}
{Hunter}, I., {Lennon}, D.~J., {Dufton}, P.~L., {et~al.} 2008, \aap, 479, 541

\bibitem[{{Iglesias} \& {Rogers}(1996)}]{1996ApJ...464..943I}
{Iglesias}, C.~A. \& {Rogers}, F.~J. 1996, \apj, 464, 943

\bibitem[{{J{\o}rgensen} \& {Lindegren}(2005)}]{2005A&A...436..127J}
{J{\o}rgensen}, B.~R. \& {Lindegren}, L. 2005, \aap, 436, 127

\bibitem[{{K{\"o}hler} {et~al.}(2014){K{\"o}hler}, {Langer}, {de Koter}, {de
  Mink}, {Sana}, {Sanyal}, \& {Schneider}}]{Koehler+2014}
{K{\"o}hler}, K., {Langer}, N., {de Koter}, A., {et~al.} 2014, \aap, submitted

\bibitem[{{Ma{\'{\i}}z Apell{\'a}niz} {et~al.}(2013){Ma{\'{\i}}z
  Apell{\'a}niz}, {Sota}, {Morrell}, {Barb{\'a}}, {Walborn}, {Alfaro}, {Gamen},
  {Arias}, \& {Gallego Calvente}}]{2013msao.confE.198M}
{Ma{\'{\i}}z Apell{\'a}niz}, J., {Sota}, A., {Morrell}, N.~I., {et~al.} 2013,
  in Massive Stars: From alpha to Omega

\bibitem[{{Ma{\'{\i}}z Apell{\'a}niz} {et~al.}(2011){Ma{\'{\i}}z
  Apell{\'a}niz}, {Sota}, {Walborn}, {Alfaro}, {Barb{\'a}}, {Morrell}, {Gamen},
  \& {Arias}}]{2011hsa6.conf..467M}
{Ma{\'{\i}}z Apell{\'a}niz}, J., {Sota}, A., {Walborn}, N.~R., {et~al.} 2011,
  in Highlights of Spanish Astrophysics VI, ed. M.~R. {Zapatero Osorio},
  J.~{Gorgas}, J.~{Ma{\'{\i}}z Apell{\'a}niz}, J.~R. {Pardo}, \& A.~{Gil de
  Paz}, 467--472

\bibitem[{{Martayan} {et~al.}(2006){Martayan}, {Fr{\'e}mat}, {Hubert},
  {Floquet}, {Zorec}, \& {Neiner}}]{2006A&A...452..273M}
{Martayan}, C., {Fr{\'e}mat}, Y., {Hubert}, A.-M., {et~al.} 2006, \aap, 452,
  273

\bibitem[{{Martayan} {et~al.}(2007){Martayan}, {Fr{\'e}mat}, {Hubert},
  {Floquet}, {Zorec}, \& {Neiner}}]{2007A&A...462..683M}
{Martayan}, C., {Fr{\'e}mat}, Y., {Hubert}, A.-M., {et~al.} 2007, \aap, 462,
  683

\bibitem[{{Penny} \& {Gies}(2009)}]{2009ApJ...700..844P}
{Penny}, L.~R. \& {Gies}, D.~R. 2009, \apj, 700, 844

\bibitem[{{Petit} \& {Wade}(2012)}]{2012MNRAS.420..773P}
{Petit}, V. \& {Wade}, G.~A. 2012, \mnras, 420, 773

\bibitem[{{Pols} {et~al.}(1997){Pols}, {Tout}, {Schroder}, {Eggleton}, \&
  {Manners}}]{1997MNRAS.289..869P}
{Pols}, O.~R., {Tout}, C.~A., {Schroder}, K.-P., {Eggleton}, P.~P., \&
  {Manners}, J. 1997, \mnras, 289, 869

\bibitem[{{Pont} \& {Eyer}(2004)}]{2004MNRAS.351..487P}
{Pont}, F. \& {Eyer}, L. 2004, \mnras, 351, 487

\bibitem[{{Ram{\'{\i}}rez-Agudelo} {et~al.}(2013){Ram{\'{\i}}rez-Agudelo},
  {Sim{\'o}n-D{\'{\i}}az}, {Sana}, {de Koter}, {Sab{\'{\i}}n-Sanjul{\'{\i}}an},
  {de Mink}, {Dufton}, {Gr{\"a}fener}, {Evans}, {Herrero}, {Langer}, {Lennon},
  {Ma{\'{\i}}z Apell{\'a}niz}, {Markova}, {Najarro}, {Puls}, {Taylor}, \&
  {Vink}}]{2013A&A...560A..29R}
{Ram{\'{\i}}rez-Agudelo}, O.~H., {Sim{\'o}n-D{\'{\i}}az}, S., {Sana}, H.,
  {et~al.} 2013, \aap, 560, A29

\bibitem[{{Salpeter}(1955)}]{1955ApJ...121..161S}
{Salpeter}, E.~E. 1955, \apj, 121, 161

\bibitem[{{Sch{\"o}nrich} \& {Bergemann}(2014)}]{Schoenrich01092014}
{Sch{\"o}nrich}, R. \& {Bergemann}, M. 2014, \mnras, 443, 698

\bibitem[{{Schroder} {et~al.}(1997){Schroder}, {Pols}, \&
  {Eggleton}}]{1997MNRAS.285..696S}
{Schroder}, K.-P., {Pols}, O.~R., \& {Eggleton}, P.~P. 1997, \mnras, 285, 696

\bibitem[{{Serenelli} {et~al.}(2013){Serenelli}, {Bergemann}, {Ruchti}, \&
  {Casagrande}}]{2013MNRAS.429.3645S}
{Serenelli}, A.~M., {Bergemann}, M., {Ruchti}, G., \& {Casagrande}, L. 2013,
  \mnras, 429, 3645

\bibitem[{{Shkedy} {et~al.}(2007){Shkedy}, {Decin}, {Molenberghs}, \&
  {Aerts}}]{2007MNRAS.377..120S}
{Shkedy}, Z., {Decin}, L., {Molenberghs}, G., \& {Aerts}, C. 2007, \mnras, 377,
  120

\bibitem[{{Sim{\'o}n-D{\'{\i}}az} {et~al.}(2011){Sim{\'o}n-D{\'{\i}}az},
  {Castro}, {Garcia}, {Herrero}, \& {Markova}}]{2011BSRSL..80..514S}
{Sim{\'o}n-D{\'{\i}}az}, S., {Castro}, N., {Garcia}, M., {Herrero}, A., \&
  {Markova}, N. 2011, Bulletin de la Societe Royale des Sciences de Liege, 80,
  514

\bibitem[{{Sim{\'o}n-D{\'{\i}}az} \& {Herrero}(2014)}]{2014A&A...562A.135S}
{Sim{\'o}n-D{\'{\i}}az}, S. \& {Herrero}, A. 2014, \aap, 562, A135

\bibitem[{{Steinmetz} {et~al.}(2006){Steinmetz}, {Zwitter}, {Siebert},
  {Watson}, {Freeman}, {Munari}, {Campbell}, {Williams}, {Seabroke}, {Wyse},
  {Parker}, {Bienaym{\'e}}, {Roeser}, {Gibson}, {Gilmore}, {Grebel}, {Helmi},
  {Navarro}, {Burton}, {Cass}, {Dawe}, {Fiegert}, {Hartley}, {Russell},
  {Saunders}, {Enke}, {Bailin}, {Binney}, {Bland-Hawthorn}, {Boeche}, {Dehnen},
  {Eisenstein}, {Evans}, {Fiorucci}, {Fulbright}, {Gerhard}, {Jauregi}, {Kelz},
  {Mijovi{\'c}}, {Minchev}, {Parmentier}, {Pe{\~n}arrubia}, {Quillen}, {Read},
  {Ruchti}, {Scholz}, {Siviero}, {Smith}, {Sordo}, {Veltz}, {Vidrih}, {von
  Berlepsch}, {Boyle}, \& {Schilbach}}]{2006AJ....132.1645S}
{Steinmetz}, M., {Zwitter}, T., {Siebert}, A., {et~al.} 2006, \aj, 132, 1645

\bibitem[{{Takeda} {et~al.}(2007){Takeda}, {Ford}, {Sills}, {Rasio}, {Fischer},
  \& {Valenti}}]{2007ApJS..168..297T}
{Takeda}, G., {Ford}, E.~B., {Sills}, A., {et~al.} 2007, \apjs, 168, 297

\bibitem[{{Torres} {et~al.}(2010){Torres}, {Andersen}, \&
  {Gim{\'e}nez}}]{2010A&ARv..18...67T}
{Torres}, G., {Andersen}, J., \& {Gim{\'e}nez}, A. 2010, \aapr, 18, 67

\bibitem[{{Torres} {et~al.}(2014){Torres}, {Vaz}, {Sandberg Lacy}, \&
  {Claret}}]{2014AJ....147...36T}
{Torres}, G., {Vaz}, L.~P.~R., {Sandberg Lacy}, C.~H., \& {Claret}, A. 2014,
  \aj, 147, 36

\bibitem[{{van Dyk} {et~al.}(2009){van Dyk}, {Degennaro}, {Stein}, {Jefferys},
  \& {von Hippel}}]{2009AnApS...3..117V}
{van Dyk}, D.~A., {Degennaro}, S., {Stein}, N., {Jefferys}, W.~H., \& {von
  Hippel}, T. 2009, Annals of Applied Statistics, 3, 117

\bibitem[{{von Hippel} {et~al.}(2006){von Hippel}, {Jefferys}, {Scott},
  {Stein}, {Winget}, {De Gennaro}, {Dam}, \& {Jeffery}}]{2006ApJ...645.1436V}
{von Hippel}, T., {Jefferys}, W.~H., {Scott}, J., {et~al.} 2006, \apj, 645,
  1436

\bibitem[{{Walmswell} {et~al.}(2013){Walmswell}, {Eldridge}, {Brewer}, \&
  {Tout}}]{2013MNRAS.435.2171W}
{Walmswell}, J.~J., {Eldridge}, J.~J., {Brewer}, B.~J., \& {Tout}, C.~A. 2013,
  \mnras, 435, 2171

\bibitem[{{Weisz} {et~al.}(2013){Weisz}, {Fouesneau}, {Hogg}, {Rix}, {Dolphin},
  {Dalcanton}, {Foreman-Mackey}, {Lang}, {Johnson}, {Beerman}, {Bell},
  {Gordon}, {Gouliermis}, {Kalirai}, {Skillman}, \&
  {Williams}}]{2013ApJ...762..123W}
{Weisz}, D.~R., {Fouesneau}, M., {Hogg}, D.~W., {et~al.} 2013, \apj, 762, 123

\bibitem[{{Yanny} {et~al.}(2009){Yanny}, {Rockosi}, {Newberg}, {Knapp},
  {Adelman-McCarthy}, {Alcorn}, {Allam}, {Allende Prieto}, {An}, {Anderson},
  {Anderson}, {Bailer-Jones}, {Bastian}, {Beers}, {Bell}, {Belokurov},
  {Bizyaev}, {Blythe}, {Bochanski}, {Boroski}, {Brinchmann}, {Brinkmann},
  {Brewington}, {Carey}, {Cudworth}, {Evans}, {Evans}, {Gates}, {G{\"a}nsicke},
  {Gillespie}, {Gilmore}, {Nebot Gomez-Moran}, {Grebel}, {Greenwell}, {Gunn},
  {Jordan}, {Jordan}, {Harding}, {Harris}, {Hendry}, {Holder}, {Ivans},
  {Ivezi{\v c}}, {Jester}, {Johnson}, {Kent}, {Kleinman}, {Kniazev},
  {Krzesinski}, {Kron}, {Kuropatkin}, {Lebedeva}, {Lee}, {French Leger},
  {L{\'e}pine}, {Levine}, {Lin}, {Long}, {Loomis}, {Lupton}, {Malanushenko},
  {Malanushenko}, {Margon}, {Martinez-Delgado}, {McGehee}, {Monet}, {Morrison},
  {Munn}, {Neilsen}, {Nitta}, {Norris}, {Oravetz}, {Owen}, {Padmanabhan},
  {Pan}, {Peterson}, {Pier}, {Platson}, {Re Fiorentin}, {Richards}, {Rix},
  {Schlegel}, {Schneider}, {Schreiber}, {Schwope}, {Sibley}, {Simmons},
  {Snedden}, {Allyn Smith}, {Stark}, {Stauffer}, {Steinmetz}, {Stoughton},
  {SubbaRao}, {Szalay}, {Szkody}, {Thakar}, {Sivarani}, {Tucker}, {Uomoto},
  {Vanden Berk}, {Vidrih}, {Wadadekar}, {Watters}, {Wilhelm}, {Wyse}, {Yarger},
  \& {Zucker}}]{2009AJ....137.4377Y}
{Yanny}, B., {Rockosi}, C., {Newberg}, H.~J., {et~al.} 2009, \aj, 137, 4377

\end{thebibliography}

\end{document}